\documentclass[]{interact}
\usepackage{amsthm}
\usepackage{graphicx}
\usepackage{color}
\usepackage{graphics,graphicx,epsfig,subfigure}
\usepackage{latexsym}
\usepackage{amssymb}
\usepackage{mathrsfs}
\usepackage{amsmath}
\usepackage{float}
\usepackage{psfrag}
\usepackage{rotating}
\usepackage{longtable}
\usepackage{supertabular}
\usepackage[misc]{ifsym}
\usepackage{gensymb}
\usepackage{adjustbox}
\usepackage[round]{natbib}
\usepackage{hyperref,xcolor}

\fontsize{11}{10}\selectfont

\begin{document}


\title{Suitability Analysis of Ground Motion Prediction Equations
	for Western and Central Himalayas and Indo-Gangetic Plains}

\author{
\name{ S. Selvan and
	Suman Sinha \thanks{Corresponding author: Suman Sinha \Letter. Email:
suman.sinha.phys@gmail.com}}
\affil{Engineering Seismology Division,
Central Water and Power Research Station, Pune, India}
}

\maketitle

\begin{abstract}
	Ground motion prediction equations (GMPEs) play a key role
	in seismic hazard assessment (SHA). Considering the seismo-tectonic,
	geophysical and geotectonic characteristics of a target region,
	all the GMPEs may not be suitable in predicting the observed ground
	motion effectively. With a fairly large number of published GMPEs, 
	the selection and ranking of suitable GMPEs for the design of 
	logic trees in SHA for a particular target region has become a 
	necessity of late. This paper presents a detailed 
	quantitative evaluation of
	performance of 16 GMPEs against recorded ground motion data in
	two target regions, characterized by distinct seismo-tectonic,
	geophysical and geotectonical nature. The data set comprises of
	465 three-component spectral accelerograms corresponding to 122 
	earthquake events. The suitability of a GMPE
	is tested by two widely accepted data-driven statistical methods,
	namely, likelihood (LH) and log-likelihood (LLH) method. Different
	suites of GMPEs are shown suitable for different periods of interest.
	The results will be useful to scientists and engineers for 
	microzonation and estimation of seismic design parameters for the design 
	of earthquake-resistant structures in these regions.
\end{abstract}

\fontsize{11}{14}\selectfont

\section{Introduction}
\label{intro}
Seismic hazard assessment (SHA) or more specifically probabilistic seismic hazard
assessment (PSHA) is the most important tool for design of earthquake resistant
structures and seismic risk mitigation. The aim of PSHA is to estimate the 
annual probability of exceedance (PoE) of a ground motion parameter of interest
for a particular target region which exhibits a characteristic attenuation of
that ground motion parameter as a function of a number of independent 
variables. For engineering purposes, the ground motion parameters are generally
expressed quantitatively in terms of peak ground acceleration (PGA) and 5$\%$
damped pseudo-spectral acceleration (PSA) at different time periods of 
engineering interest. The PGA and PSA are predicted from ground motion
prediction equations (GMPEs) which are model fitting to 
represent the functional form of recorded data by regression analysis.
The GMPEs estimate PGA or PSA as a function of independent variables such as
earthquake magnitude, source-to-site distance and site amplification effects
along with various other parameters such as fault type, hanging wall effect 
etc. Therefore, for a reliable and realistic estimate of ground motion 
parameters, the selection of appropriate GMPEs is of extreme importance. The
selected GMPEs should represent the ground motion of the region under study
in a realistic manner. It needs to be mentioned here that the prediction of 
ground motion parameters from GMPEs depends on various parameters, all of 
which are not precisely known and therefore cannot be taken into account with
confidence. Therefore, certain degree of subjectivity is always associated
with the implementation of GMPEs. These uncertainties are classified into 
two types: (i) aleatory uncertainty which is taken care of by the residual
scatter, generally denoted by $\sigma$, of the GMPEs and (ii) 
epistemic uncertainty which arises due to insufficient knowledge.
For example, a number of GMPEs may be proposed for a particular target 
region and there is no \textit {a priori} justification that one is more 
acceptable
than the others (which indicates the lack of knowledge) 
\citep{cott1}.
This epistemic 
uncertainty is taken care of by involving multiple GMPEs in a logic tree 
framework \citep{kulk1,bomm2}. Each branch of the logic tree corresponds to a different
GMPE and proper weights should be assigned to each branch (\textit{i. e.}
each GMPE) in the PSHA calculations to arrive at the final value of the
ground motion parameters. In other words, 
data-driven selection and ranking of the GMPEs
for performing SHA in a target region is necessary.

Different quantitative data-driven methods \citep{sche1,sche2,kale1,kowa1}
have been proposed so far to guide the selection of appropriate GMPEs. Among
the existing methods, \citet{sche1} first proposed a data-driven method, 
based on
exceedance probabilities, to quantify the suitability of a candidate GMPE 
against recorded data. The method, now widely known as likelihood (LH) method,
although not free from certain subjectivities, is based on sound
mathematical background. Later \citet{sche2} proposed another data-driven 
quantitative method, based on information theory, to decide the appropriateness
of a candidate GMPE and thereafter assigning the weights to the branch of the
logic tree. This method, now widely known as log-likelihood (LLH) method, does 
not depend on \textit {ad hoc} assumptions \citep{dela1} and therefore 
overcomes the limitations of the LH method.

The suitability of GMPEs in different regions has earlier been tested
\citep{hint1,staf1,aran1,beau1,byko1,sum22} using different methods. 
Here the suitability of various GMPEs
are examined by performing a thorough quantitative assessment in a
systematic way to target regions which exhibit complex seismo-tectonic setup
and are characterized by distinct geophysical and geotechnical properties. A
detailed systematic study of applicability of the GMPEs for the said target
regions has not been carried out earlier to the best of our knowledge.

The following section gives a brief description of target regions, followed
by a description of data sets used in the study and then list of GMPEs
considered for the present study. Section \ref{ddm} elucidates
the data-driven methods. The next section presents the results in each
target region with detailed discussions of the results. Section \ref{conclu} 
draws the conclusions.

\section{Target Regions with their Seismotectonic Framework}
\label{ros}
\begin{figure}[!h]
\begin{center}
        \rotatebox{0}{\includegraphics[scale=0.65]{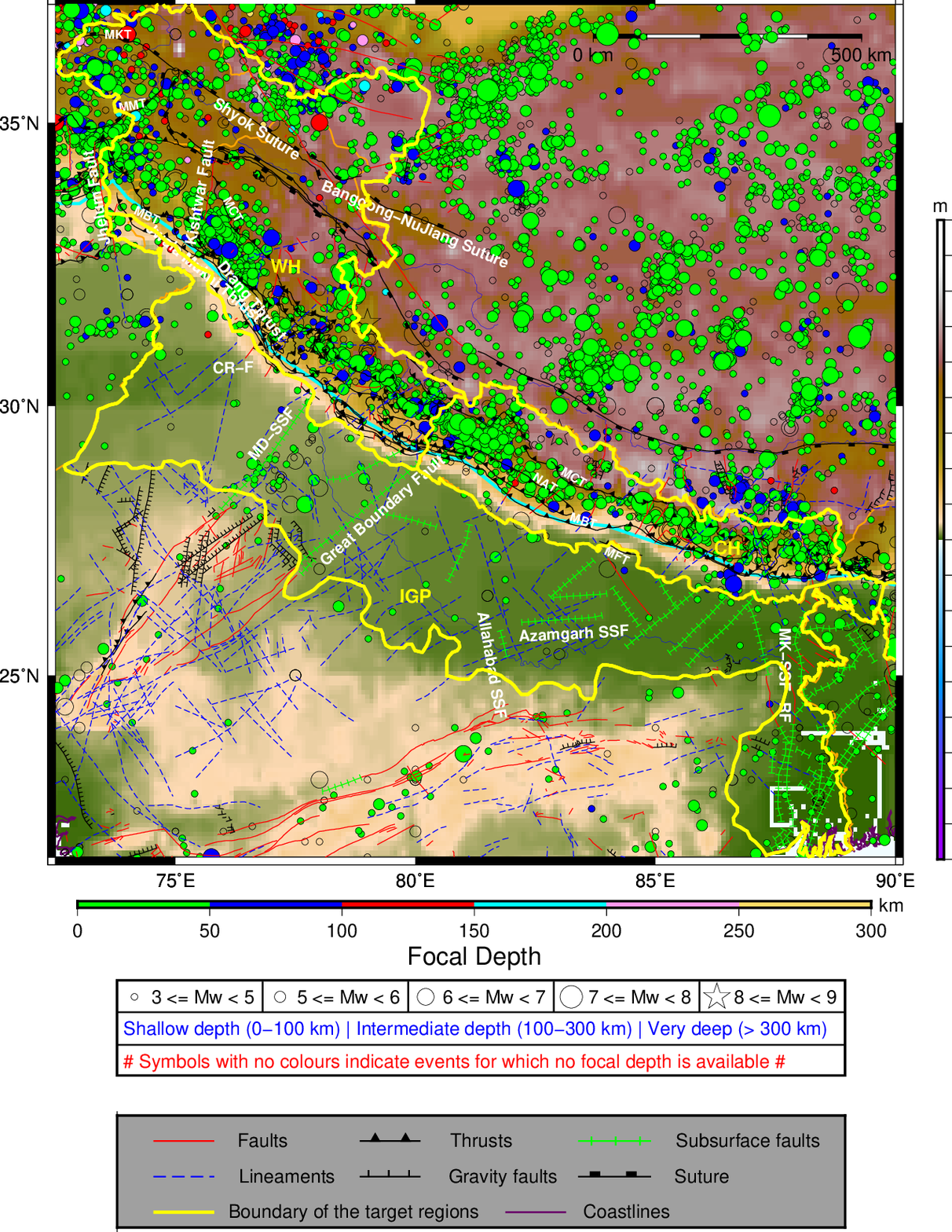}}
	\caption{Map showing the target regions with various tectonic features
	\citep{dasg1}
	and the epicenters of past earthquakes superimposed therein.
	The international boundaries and coastlines are produced from
	the Generic Mapping Tools (GMT) database. WH stands for western
	Himalaya, CH stands for central Himalaya and IGP stands for 
	Indo-Gangetic Plains.}
\label{tr}
\end{center}
\end{figure}

Knowledge of seismotectonic characteristics of a region is important
to ascertain the seismic potential of the region and to evaluate the
probable ground motion that it can generate. The two target regions,
considered for the present study,
constitute intricate and complex seismotectonic environment and 
exhibit distinct geological characteristics. These two target regions, 
namely, (i) Western and Central Himalayas and 
(ii) Indo-Gangetic Plains are shown in Fig. \ref{tr}. 
The epicenters of past earthquakes, taken from
the reviewed  International Seismological
Centre (ISC) bulletin, UK
(\url{http://www.isc.ac.uk/iscbulletin/search/bulletin/}),
National Earthquake Information Centre (NEIC),
(\url{https://earthquake.usgs.gov/earthquakes/search/}), USGS and
India Meteorological Department (IMD), New Delhi and various tectonic features
taken from \citep{dasg1} are also plotted in Fig. \ref{tr}.
\citet{gup06} has extensively
analysed the seismotectonic characteristics of the Himalayas and
northeast India and brought out a wide picture of seismic sources in
these regions. 

\citet{kaya22}, after analysing
the seismic cross sections and focal plane solutions of the region,
suggested that the source process of the
entire Himalayas have heterogeneous seismotectonic patterns.
\subsection{The Western and Central Himalayas}
The western and central Himalayas within the framework of the
Himalayan arc tectonic belt is dominated by seismically
active thrust faults which produce large magnitude earthquakes
due to the interaction between Indian and Eurasian continental
plates. The geology of the entire region is possessed by various rock
formations such as schist, quartzite, slates, phyllites, gneisses and granites.
The predominant thrust faults include Main Central Thrust (MCT),
Main Boundary Thrust (MBT) and Main Frontal Thrust (MFT) along with many
subsidiary thrusts. The decollement surface which is the plane of
detachment between Indian and Eurasian plates is considered to be
responsible for few of the great earthquakes (1991 Uttarkashi and
the 1999 Chamoli earthquakes having magnitude $>$ 6.0) that occurred
north of MBT in this region \citep{kaya10}.
The 2015 Nepal earthquake of $M_w$ 7.8 was caused by reverse faulting
at the down-dip portion of the Main Himalayan Thrust (MHT) \citep{ajay1}.
Apart from the thrust
faults, many
transverse and oblique faults also occupy this
region which cause the long extends of MBT and MCT break into fragments
\citep{gup06}. \citet{kaya14}  reports few great earthquakes caused
by strike-slip faulting in this region (1988 Bihar/Nepal foothill Himalaya
and 2011 Sikkim Himalaya earthquakes having magnitude $>$ 6.0).
\citet{mona1} reported a maximum magnitude 7.7 for western Himalaya and
8.0 for central Himalaya.
\subsection{The Indo-Gangetic Plains}
The Indo-Gangetic plains are large floodplains of the Indus, Ganga, Yamuna
and the Brahmaputra river systems \citep{nar21}. They run parallel
to the Himalaya mountains, from Jammu and Kashmir and Khyber Pakhtunkhwa
in the west to the western part of northeast India in the east. This region,
even though characterised by moderate earthquakes, has high influence
of seismic activities in the adjacent western and central Himalayas and shows
widespread alluvial deposits which constitute considerable site amplifications.
This region witnesses several subsurface faults (SSFs) including Azamgarh SSF
and Allahabad SSF. The Indus basin is comprised of Chandhigar Ropar fault
(a neotectonic fault), North-Delhi fold belt and Mahendragarh and Dehradun
SSFs. The Ganga basin contributes many dip slip, oblique and strike slip faults
which cross transverse to the Himalayan Frontal thrust zone. The quaternary
alluvial deposits in the Bengal basin cover the state of West Bengal.
Maldah-Kishanganj fault, Rajmahal fault and many
neotectonic faults are present in the Bengal basin with moderate seismic
activities \citep{dasg1}.
\citet{anba3} reported a maximum magnitude of 7.0 by considering the 
regional rupture characteristics for Patna region which
lies in Indo-Gangetic plains while a maximum magnitude of 6.6 was reported
by \citet{mona1} for Indo-Gangetic plains.

\section{Ground-Motion Data Set}
\label{ds}
A total of 122 earthquakes from 163 recording 
Stations have been considered for
for performing
the efficacy test of the GMPEs for the said target regions.
These earthquakes
correspond to a total of 465 three-component ground-motion accelerograms 
(\textit {i.e,} $465 \times 2 = 930$ horizontal accelerograms) which are
used for examining the suitability of GMPEs in the
said target regions. Out of this 465 records, 356 records ($356 \times 2 =
712$ horizontal accelerograms) from 96 recording Stations 
are from northwest and central Himalayas while
109 records ($109 \times 2 = 218$ horizontal accelerograms) from 
53 recording Stations are from Indo-Gangetic plains.
It is observed that western side of IGP has good distribution of 
recording stations compared to the eastern side of IGP. However, out of 53 stations
in IGP, Delhi and its surrounding area have 32 stations while the remaining 21
stations are distributed in the rest of IGP.

\begin{figure}[!h]
\begin{center}
        \rotatebox{0}{\includegraphics[scale=0.65]{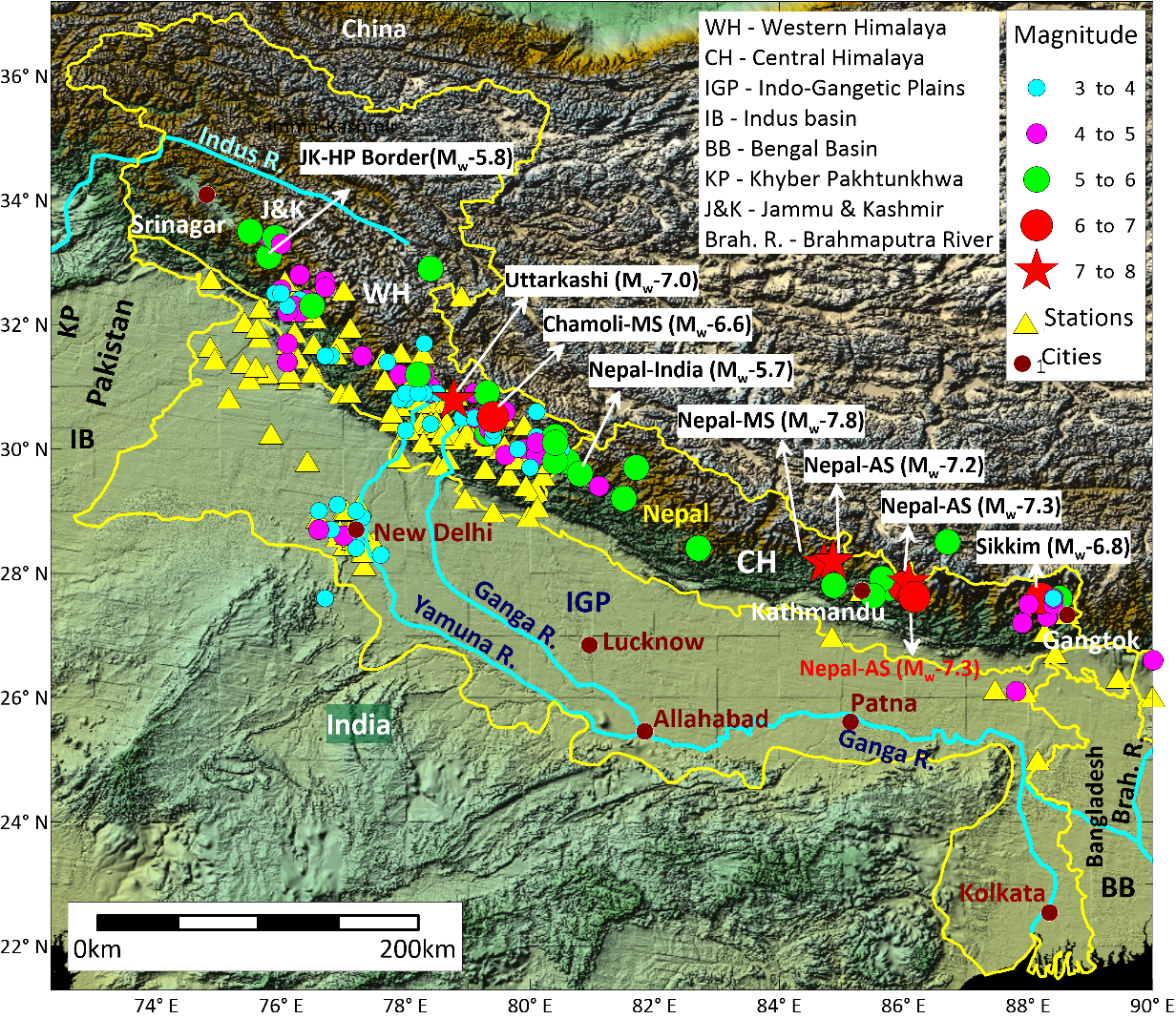}}
	\caption{Map showing the locations of recording stations and 
	earthquake epicenters of the
	recorded data used in the study. The boundary of the two target
	regions are marked in yellow color. The DEM data are taken from
	GLOBE model of NOAA.}
\label{eqrs}
\end{center}
\end{figure}

\begin{figure}[!h]
\begin{center}
        \rotatebox{0}{\includegraphics[scale=0.55]{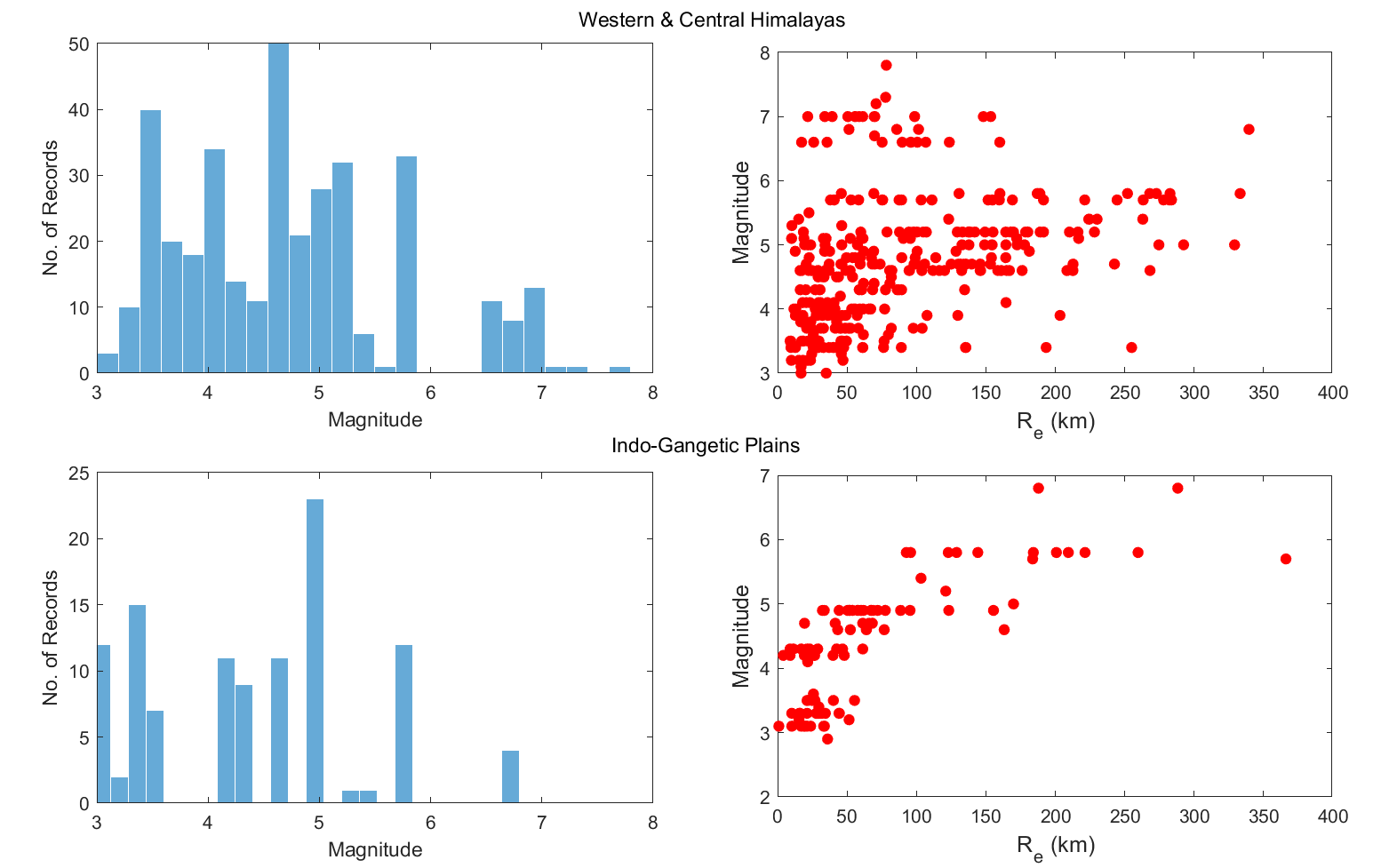}}
	\caption{Distribution of earthquake records (left panel) and 
	magnitude-distance scatter plot (right panel) of the data set used
	in the study.}
\label{dataset}
\end{center}
\end{figure}

For northwest and central Himalaya region,
the magnitude and the distance (epicentral) range of the earthquake events are
3.0 - 7.8 and 2 - 340 km respectively and the same for Indo-Gangetic plains
are 3.0 - 6.8 and 1 - 366 km respectively. 
The locations of the earthquake epicenters and the 
recording stations are shown in Fig. \ref{eqrs}.
The region under study has witnessed several devastating earthquakes in the
past. Some significant earthquakes whose acclerograms are used in the 
present analysis are listed in Table \ref{eqtab}.
\begin{table}[!h]
        \caption{Some significant earthquakes in the 
        region of study}
\label{eqtab}
\begin{adjustbox}{max width=1.1\textwidth,center}
\begin{tabular}{lcccccl}
\hline\noalign{\smallskip}
	Name of  & Date & Lat & Lon & Focal Depth & Magnitude & Fault  \\
	 Earthquake  & (dd-mm-yy) & ($\degree N$) & ($\degree E$) & (km) & ($M_W$) & Type\\
\noalign{\smallskip}\hline\noalign{\smallskip}
Chamoli-Mainshock	& 28-03-1999 &	30.512	& 79.403 & 15.0	&	6.6	& Reverse \\
Chamoli-Aftershock	& 28-03-1999 &	30.315	& 79.380 & 10.0	&	5.4	& Reverse \\
Chamoli-Aftershock	& 30-03-1999 &	30.376	& 79.330 & 10.0	&	5.3	& Reverse	\\
Chamoli-Aftershock	& 06-04-1999 &	30.414	& 79.320 & 10.0	&	5.1	& Reverse	\\
Chamoli-Aftershock	& 07-04-1999 &	30.250	& 79.320 & 10.0	&	5.0	& Reverse	\\
Chamoli-Aftershock	& 07-04-1999 &	30.260	& 79.310 & 10.0	&	5.2	& Reverse	\\
JK-HP-Border		& 01-05-2013 &	33.100	& 75.800 & 15.0	&	5.8	& Strike-slip	\\
Nepal-India Border	& 04-04-2011 &	29.600	& 80.800 & 10.0	&	5.7	& Strike-slip	\\
India(Sikkim)-Nepal	& 18-09-2011 &	27.600	& 88.200 & 10.0	&	6.8	& Strike-slip	\\
India(Sikkim)-Nepal (AS)& 18-09-2011 &	27.600	& 88.500 & 16.0	&	5.0	& Strike-slip	\\
Uttarkashi		& 19-10-1991 &	30.780	& 78.774 & 10.0	&	7.0	&	Reverse	\\
Nepal-Mainshock		& 25-04-2015 &	28.150	& 84.710 & 15.0	&	7.8	&	Reverse	\\
Nepal-Aftershock	& 26-04-2015 &	27.760	& 85.770 & 10.0	&	5.3	&	Reverse	\\
Nepal-Aftershock	& 26-04-2015 &	27.780	& 86.000 & 17.3	&	6.7	&	Reverse	\\
Nepal-Aftershock	& 25-04-2015 &	27.910	& 85.650 & 10.0	&	5.7	&	Reverse	\\
Nepal-Aftershock	& 25-04-2015 &	28.190	& 84.870 & 14.6	&	7.2	&	Reverse	\\
Nepal-Aftershock	& 25-04-2015 &	27.805	& 84.874 & 10.0	&	5.8	&	Reverse	\\
Nepal-Aftershock	& 25-04-2015 &	27.640	& 85.500 & 10.0	&	5.5	&	Reverse	\\
Nepal-Aftershock	& 12-05-2015 &	27.840	& 86.080 & 15.0	&	7.3	&	Reverse	\\
Nepal-Aftershock	& 12-05-2015 &	27.620	& 86.170 & 15.0	&	6.8	&	Reverse	\\
\noalign{\smallskip}\hline
\end{tabular}
\end{adjustbox}
\end{table}
The distribution of dataset in terms of magnitude and
distance is presented in Fig. \ref{dataset}.
The variation in the range of magnitude and distance in
the data set make it appropriate to perform the efficacy test for different
GMPEs. 
It is mentioned here that there are 21 aftershocks out of
a total of 122 earthquakes and 41 earthquakes having magnitude lying in the 
range 3 $\le$ $M_w$ $<$ 4 are there.
	Magnitudes below 4.0 are used to test the suitability of GMPEs as
NGA-West2 GMPEs are developed by considering earthquakes having 
magnitudes less than 4.0.
Although magnitudes less than 4.0 should not contribute significantly 
to seismic hazard estimation, \citet{slej1} reported that,
in moderate seismicity region, the 
lower threshold value of magnitude can be taken as 3.0.

It is worth mentioning that the data set considered
for the present study was mostly not used by the developers of the GMPEs and
is an independent data set.
The ground-motion data set, used for the present study has been
downloaded from PESMOS (\url {https://pesmos.org}). In addition, the
PESMOS data have been supplemented by $53$ three-component ground-motion
accelerograms (a total of $53 \times 2 = 106$ horizontal accelerograms)
from $21$ earthquakes of different hypo-central depths and these $53$
records have been downloaded from Center for Engineering Strong Motion Data
(\url {https://www.strongmotioncenter.org}). 
The ground motion data set downloaded from the above sources
are processed only (baseline corrected and low pass filtered).
The processed downloaded data have been compiled in a consistent
manner to tabulate magnitude, source-to-site distance metrics, local
site conditions and PGA. The 5$\%$ damped PSA at different time periods
have been computed from the processed acceleration time histories using
standard computer programs.

It is added here that the site characteristics of all the 
recording stations were provided in the data set downloaded. While computing
the PGA or PSA (at different time periods), the same site conditions were
used to avoid any bias in the selection of suitable GMPEs. The set of suitable
GMPEs, thus selected, may be used for computation of PGA or PSA 
(at different time periods) at any site inside a target region. The site 
condition of that particular site will then be an input parameter in the 
selected GMPEs. The details of the recording stations are given in the 
Supplementary Material.

\section{List of GMPEs Considered}
There have been considerable advances in the development of GMPEs with the
increase of seismic networks since 2000. A very good account of the
empirical GMPEs existing throughout the world for the period 1964-2021
is well documented in the report by \citet{doug1}. Following the minimum 
criteria proposed by \citet{cott1} and \citet{bomm1}, 16 GMPEs for crustal
earthquakes have been primarily
selected for performing the efficacy test and ranking of the GMPEs.
All the GMPEs are mostly recent GMPEs, robust and adequately constrained.
Out of the 16 GMPEs, 12 GMPEs have been developed from world wide data and
the remaining are developed for other regions including one for Himalayan
region. All the GMPEs with their
range of applicability is listed in Table \ref{gmpetab}.

It needs to be mentioned here that limited seismic network and paucity of
ground motion data in India imposes a limitation for developing robust
GMPEs empirically, exclusively for the regions under study. It is also added 
here that the observation of \citet{cott1} that a GMPE, developed for a region
which does not conform tectonically to the region under study should be
excluded, cannot possibly be true as the selection or exclusion of a GMPE
can never be based on geographical criteria \citep{byko1,bomm1}. Several
studies show that there is no solid evidence of regional differences in
ground motions in tectonically comparable regions, at least in the range of
moderate to high magnitude earthquakes \citep{staf1,doug2}, with few 
exceptions in active regions \citep{stra1} for high-frequency response
spectra. Rather, this criterion should be interpreted as to exclude the
GMPEs, developed for subduction earthquakes, in PSHA calculations for shallow
crustal earthquakes and vice versa, or for example, the GMPEs made for
volcanic areas should not be used in regions which does not exhibit volcanic
activities. Therefore, after preliminary analysis, the GMPEs, listed in
Table \ref{gmpetab}, have been considered for the target regions.

It is mentioned here that few region-specific GMPEs 
\citep{shar1, nath3, singh1, gupta5, hari1} are not shown in the present analysis for
practical limitations with these GMPEs in the context of the present
analysis. For example, \citet{shar1} and \citet{gupta5} considered time
periods starting from 0.04 s. Now, the PSA at 0.04 sec cannot be treated 
equivalent to PGA. 
On the other hand, \citet{nath3} proposed their GMPE for 8 periods 
(including PGV) only ranging from PGA to 4.0 s with wide intervals. 
The PSA at many periods in between are not 
available, which makes it difficult to arrive at a conclusive result.
The GMPE, proposed by \citet{singh1} considered only one main 
shock (Nepal earthquake) and five aftershocks.
The database is fairly small for developing a robust GMPE. This GMPE did 
neither consider the fault type nor the soil response term. Moreover, 
0.2 s time period, which is a crucial one for this study, is not there in this GMPE. 
The GMPE, proposed by \citet{hari1}, considered a small database 
(only 15 earthquakes), of which 4 events are above 6.0 $M_w$. 
Due to these practical limitations, these regional 
GMPEs were not found suitable for testing the LH and the LLH methods
as the LH ranks and LLH scores are expected
to be inconsistent compared with the other GMPEs considered. 
It is pointed out that the three most important criteria for 
primarily rejecting a GMPE for consideration, as per \citet{cott1}, are
    (i) The underlying dataset for developing the GMPE is insufficient,
    (ii) The frequency range of the model is not appropriate for engineering application
and (iii) The model has an inappropriate functional form.
The present analysis is focussed on testing two well established methods 
for selection of best suitable GMPEs in two target regions and GMPEs that 
allow to arrive at a conclusive result are considered.

\begin{table}[!h]
	\caption{GMPEs considered in this study with their range 
	of applicability} 
\label{gmpetab}
\begin{adjustbox}{max width=1.2\textwidth,center}
\begin{tabular}{lccccl}
\hline\noalign{\smallskip}
	GMPEs with abbreviations & Magnitude & Distance & $V_{s30}$ & Periods & Region   \\
	and references & ($M_W$) range & range (km) & range (m/s) & (s) & \\
\noalign{\smallskip}\hline\noalign{\smallskip}
	KAN06 (\citet{kann1}) & 5.5 - 8.2 & $<$ 450 & 150 - 1500 & 0.0 - 5.0 & Japan \\ 
	ZHAO06 (\citet{zhao06})	&	5 - 7.5	&	0 - 300	&	4 site classes	&	0.0 - 5.0	&	Japan \\
	AS08 (\citet{abra2}) & 5 - 8.5  & $<$ 200 & $>$ 180 & 0.0 - 10 & Worldwide \\
	BA08 (\citet{boore1}) & 5 - 8.0 & 0 - 200 & 180 - 1300 & 0.0 - 10 & Worldwide \\
	CB08 (\citet{cam4}) & 4 - 8.5 & 0 - 200 & 150 - 1500 & 0.0 - 10 & Worldwide \\
	CY08 (\citet{chi1}) & 4 - 8.5 & 0 - 200 & 150 -1500 & 0.0 - 10 & Worldwide \\
	IDR08 (\citet{idr08})	&	4.5 - 8	&	0 - 200	& $>$ 450	&	0.0 - 10	&	World wide \\
	AKBO10 (\citet{akbo10})	& 5 - 7.6 & $<$ 100 & 3 site classes & 0 - 3.0 & EMM \\
	ASK14 (\citet{ask14})	&	3 - 8.5	&	0 - 300	&	180 - 1500	&	0.0 - 10	&	World wide \\
	BSSA14 (\citet{bssa14})	& 3.0 - 8.5 (SS, RV) 3.3 - 7.0 (NM) & 0 - 400	& 150 - 1500 &	0.0 - 10 & World wide \\
	CB14 (\citet{cb14}) & 3.3 - 8.5 (SS)     3.3 - 8.0 (RV)    3.3 - 7.0 (NM) & 0 - 300 & 150 - 1500 & 0.01 - 10 &	World wide \\
	CY14 (\citet{cy14}) & 3.5 - 8.5 (SS)    3.5 - 8.0 (RV, NM) &	0 - 300	& 180 - 1500 &	0.0 - 10 & World wide \\
	IDR14 (\citet{idr14})	&	5 - 8.0	&	0 - 150	&	450 - 2000	&	0.0 - 10	&	World wide \\
	GK15 (\citet{gk15})	&	5.0 - 8.0	&	0 - 250	&	200 - 1300	&	0.01 - 5.0	&	World wide \\
	ZHAO16 (\citet{zhao16c}) &	5 - 7.5 & 0 - 300	&	4 site classes	&	0.0 - 5.0	&	Worldwide \\
	BJAN19 (\citet{bjan19})	&	4 - 9.0	&	10 - 750	&	NA	&	0.0 - 10	&	Himalaya \\
\noalign{\smallskip}\hline\noalign{\smallskip}
	* SS = Strike slip, RV = Reverse, NM = Normal; & EMM = Europe, Mediterranean and Middle East & & & & \\
\noalign{\smallskip}\hline
\end{tabular}
\end{adjustbox}
\end{table}
\section{Data-driven methods}
\label{ddm}
In this section, we will discuss two methods, namely the LH and the LLH 
method, to judge the goodness-of-fit measures between observed ground
motion data for a target region and a candidate GMPE. There are several 
methods for evaluating the goodness-of-fit measures \citep{sche1,sche2,
kale1,kowa1}. LH method is a transparent one which is based on the
concept of likelihood \citep{edwa1} and is particularly useful in examining
the match or mismatch of recorded data to a GMPE for a target region.
The level of agreement between the observations and the predictions in
LH method is quantitatively expressed in terms of mean, median and standard
deviation of the normalized residual and the median values of the LH
parameter as well. Under the assumption that a ground motion parameter
(say $y$) obeys log-normal distribution (\textit {i.e}, $\ln y$ is normally
distributed), the normalized residual is expressed as
\begin{equation}
        Z=\frac{\ln( {y})_{\tt obs}-\ln( {y})_{\tt predicted}}
        {\sigma_{\tt T}}
\label{nr}
\end{equation}
where $( {y})_{\tt obs}$ and $({y})_{\tt predicted}$
are the observed and the predicted ground motions and
$\sigma_{\tt T}$ is the total standard deviation of the GMPE
used. So normalized residual ($Z$) represents the distance of the
data from the logarithmic mean measured in units of
$\sigma_{\tt T}$. The LH parameter is calculated as
\begin{equation}
LH(\vert Z \vert)= Erf(\frac{\vert Z \vert}{\sqrt 2}, \infty)
= \frac{2}{\sqrt{2\pi}}\int_{\vert Z \vert}^{\infty} exp\left(\frac{-Z^2}{2}\right)dZ
\label{lh}
\end{equation}
where $Erf$ is the error function.
The median of $LH$ plays an important role in quantifying the suitability of
a GMPE. If the chosen GMPE matches the observed data well in terms of both
the mean and the standard deviation, then the $LH$ values are evenly distributed
between $0$ and $1$ and the median of $LH$ is $\sim 0.5$. In case the 
GMPE is unbiased in terms of mean but the standard deviation of the observed
data is either smaller or greater than the standard deviation of the GMPE,
the distribution of $LH$ becomes asymmetric and the median value of $LH$ is
either $>0.5$ or $<0.5$. The more the standard deviation of the observed
data and the shift of the mean value, the less is the median of the $LH$
distribution. These properties make the $LH$ method a good one to check the
suitability of a GMPE against recorded data. Although, it is not only the median
value of the $LH$ distribution but also the mean, median and the standard
deviation of the normalized residual that decide the goodness-of-fit 
measures of $LH$ method. Depending on these measures, \citet{sche1} 
proposed a classification scheme for ranking of GMPEs to evaluate the
performance of GMPEs against recorded ground motion data. The four
goodness-of-fit measures for deciding the rank of a GMPE are listed in
Table \ref{rank}.
\begin{table}[!h]
\caption{Values of goodness-of-fit measures for deciding the rank of a GMPE}
\label{rank}
\begin{adjustbox}{max width=1.1\textwidth,center}
\begin{tabular}{ccccc}
\hline\noalign{\smallskip}
	$\mid \text{Mean} \ \textit{Z} \mid$ & $\mid \text{Median}\ \textit{Z}\mid$ & Std $Z$ & Med. $LH$ & Rank \\
\noalign{\smallskip}\hline\noalign{\smallskip}
$<$ 0.25 & $<$ 0.25  & $<$ 1.125  & $>$ 0.4 & A (high capability)  \\
$<$ 0.50 & $<$ 0.50  & $<$ 1.250  & $>$ 0.3 & B (medium capability)  \\
$<$ 0.75 & $<$ 0.75  & $<$ 1.500  & $>$ 0.2 & C (low capability)  \\
 & & All other combinations & & \\
 & & of goodness-of-fit measures & & D (unacceptable capability)  \\
\noalign{\smallskip}\hline
\end{tabular}
\end{adjustbox}
\end{table}
However, the $LH$ method is dependent on the sample size, \textit{i.e},
the total number of recorded grounded motion data and despite its
statistical relevance, the $LH$ method is not completely free from 
certain subjectivities. The subjectivities lie in the threshold values
of the goodness-of-fit measures and in the definition of ranks. To overcome
these limitations, \citet{sche2} proposed another goodness-of-
fit method, the so-called LLH method, which does not depend on \textit{ad hoc}
assumptions.

LLH method is based on the concepts of information theory. The fundamental
aim of LLH method is to measure the Kullback-Leibler (KL) distance 
between two models - a model that represents reality (here the ground motion
data) and a candidate ground motion model (here a GMPE). If two models
$f$ and $g$ be described by two probability density functions $f(x)$ and
$g(x)$, the KL distance between the two models is expressed as \citep{sche2}
\begin{equation}
	D(f,g) = E_{f}[\log_2 (f)] - E_{f}[\log_2 (g)]
\label{KLd}
\end{equation}
where $E_{f}$ is the statistical expectation of the log-likelihood of the
model function with respect to $f$. $D(f,g)$ is interpreted as the 
relative entropy between $f$ and $g$ \citep{cove1} and represents the amount
of information loss if model $f$ is replaced by model $g$. The negative
value of the statistical expectation of the log-likelihood function indicate
a natural distance measure \citep{sche2}. For comparison of models, the
relative KL distance is of only interest as the statistical expectation of
the log-likelihood of $f$ with respect to $f$ cancels out as a constant term
\citep{sche2}. Therefore, to specify a ranking criterion, the log-likelihood
score may be defined as 
\begin{equation}
	LLH = -\frac{1}{N} \sum_{n=i}^{N} \log_2(g(x_i))
\label{llheq}
\end{equation}
where $N$ is the total number of observations. This LLH score is termed as
average sample log-likelihood. The Eqn. \ref{llheq} for LLH score is more
explicitly written as 
\begin{equation}
	LLH = -\frac{1}{N} \sum_{n=i}^{N} \log_2 \left[\frac{1}{\sigma_i 
	\sqrt{2\pi}}\exp\left\{-\frac{1}{2}\left(\frac{y_i-\mu_i}
	{\sigma_i}\right)^2\right\}\right]
\label{llheq1}
\end{equation}
where $y_i$ is the observed ground motion value and $\mu_i$ and $\sigma_i$ 
are the mean and the standard deviation of the GMPE being considered. A
lower LLH score implies a better performance of a GMPE for a target region.
In LLH method, only one measure, \textit{i.e}, LLH score, which basically
indicates the average loss of information when the model representing reality
is replaced by a candidate GMPE, is required to examine the suitability of
a GMPE. On the contrary, a combination of four measures is required to decide
the ranking criteria in LH method, as discussed earlier. Therefore the LLH
method may be treated as a better one.
\section{Results and discussion}
\label{rd}
In this section, the results obtained from our analysis are presented.
The analyses are carried out in two target regions and the performance of
16 GMPEs is evaluated at 20 periods using the two data-driven methods,
discussed in the previous section.
\subsection{Results for Western and Central Himalaya}
\begin{figure}[!h]
\begin{center}
        \rotatebox{0}{\includegraphics[scale=0.6]{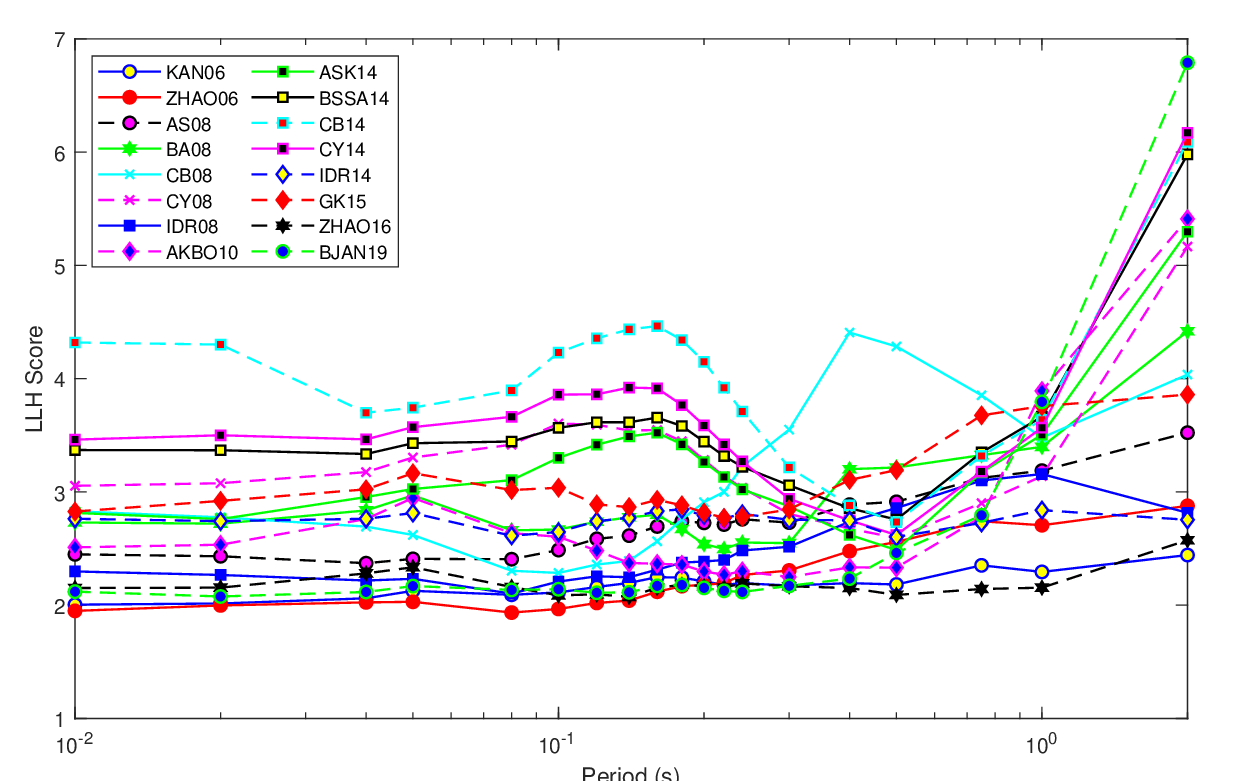}}
	\caption{Plot of LLH score at 19 periods for the all the GMPEs considered
	in western and central Himalayas.}
\label{llhper1}
\end{center}
\end{figure}

\begin{figure}[!h]
\begin{center}
        \rotatebox{0}{\includegraphics[scale=0.6]{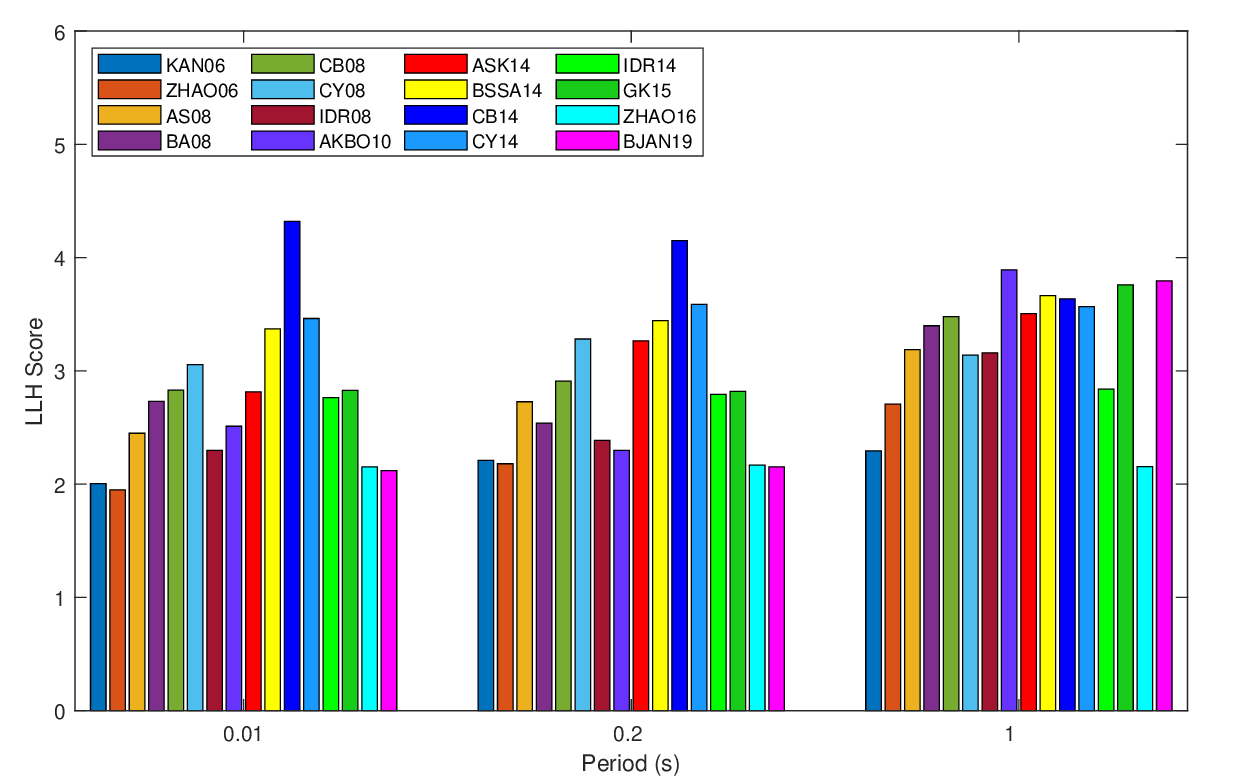}}
	\caption{Bar diagram of LLH score at $0.01$s, $0.2$s and 
	$1.0$s for the all the GMPEs considered in western and central 
	Himalaya.}
\label{llhper2}
\end{center}
\end{figure}

\begin{figure}[!h]
	\begin{adjustbox}{max width=1.1\textwidth,center}
\begin{tabular}{c}
      \resizebox{120mm}{!}{\rotatebox{0}{\includegraphics[scale=0.6]{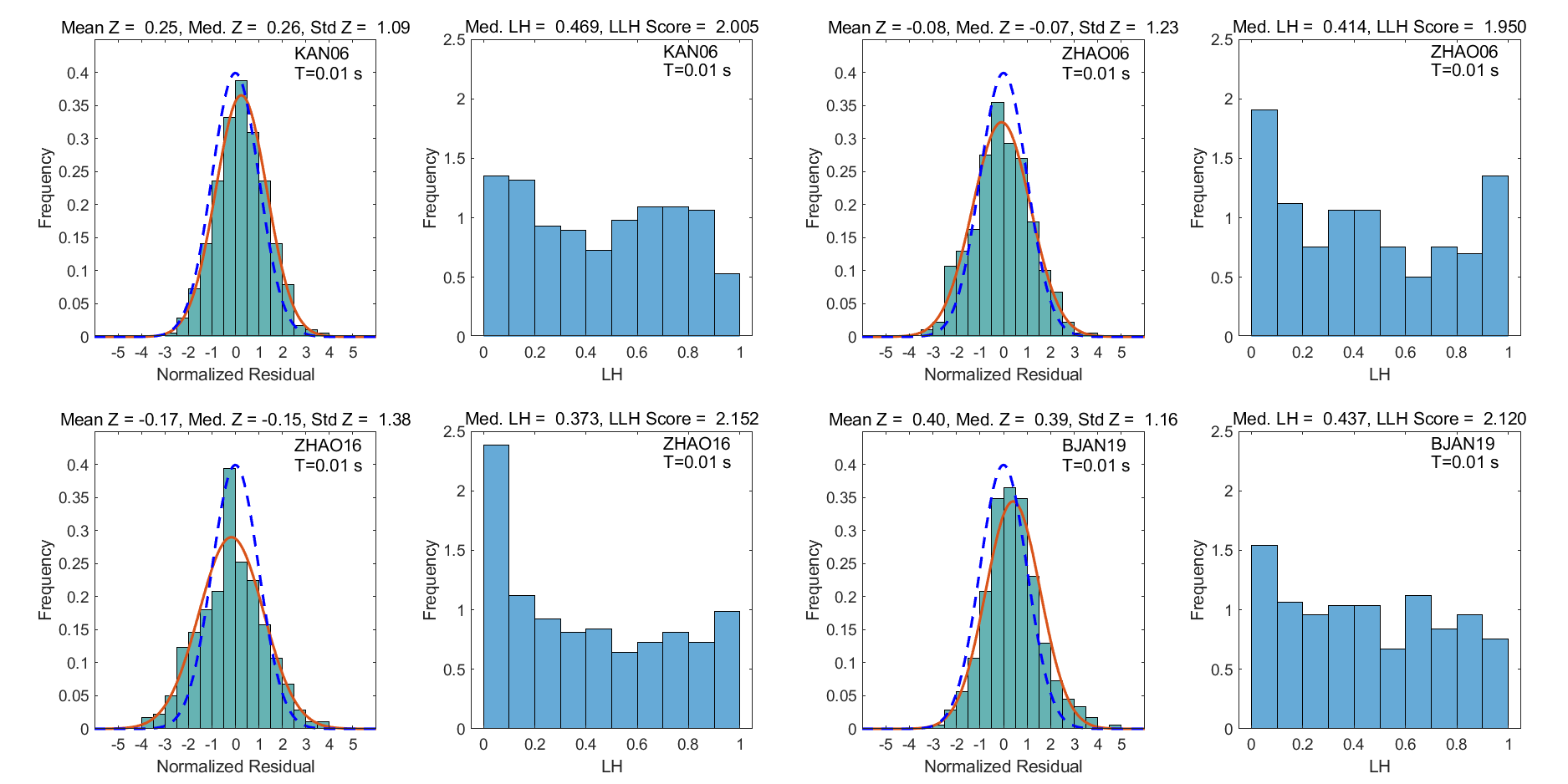}}} \\
      \resizebox{120mm}{!}{\rotatebox{0}{\includegraphics[scale=0.6]{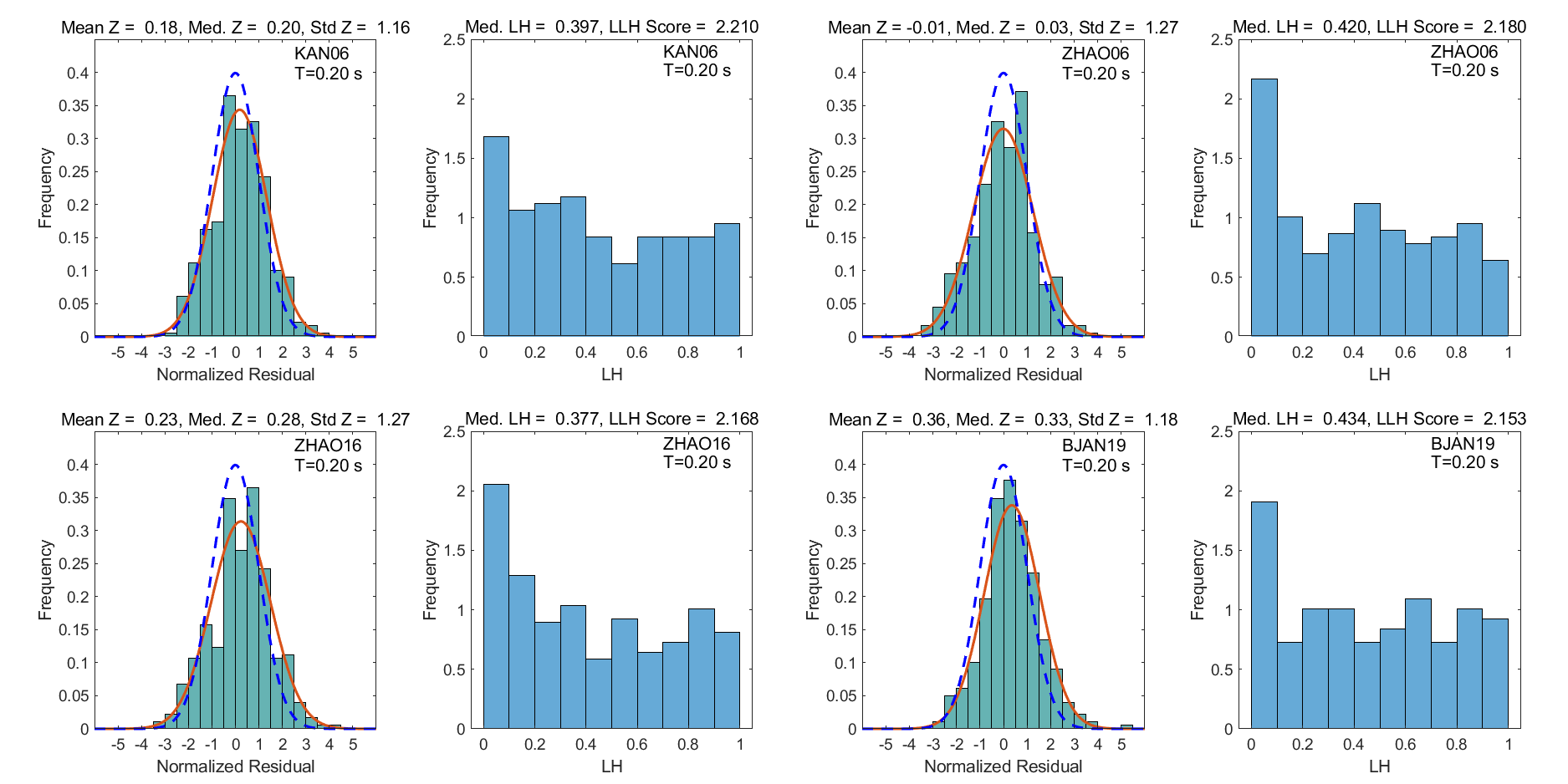}}} \\
      \resizebox{120mm}{!}{\rotatebox{0}{\includegraphics[scale=0.6]{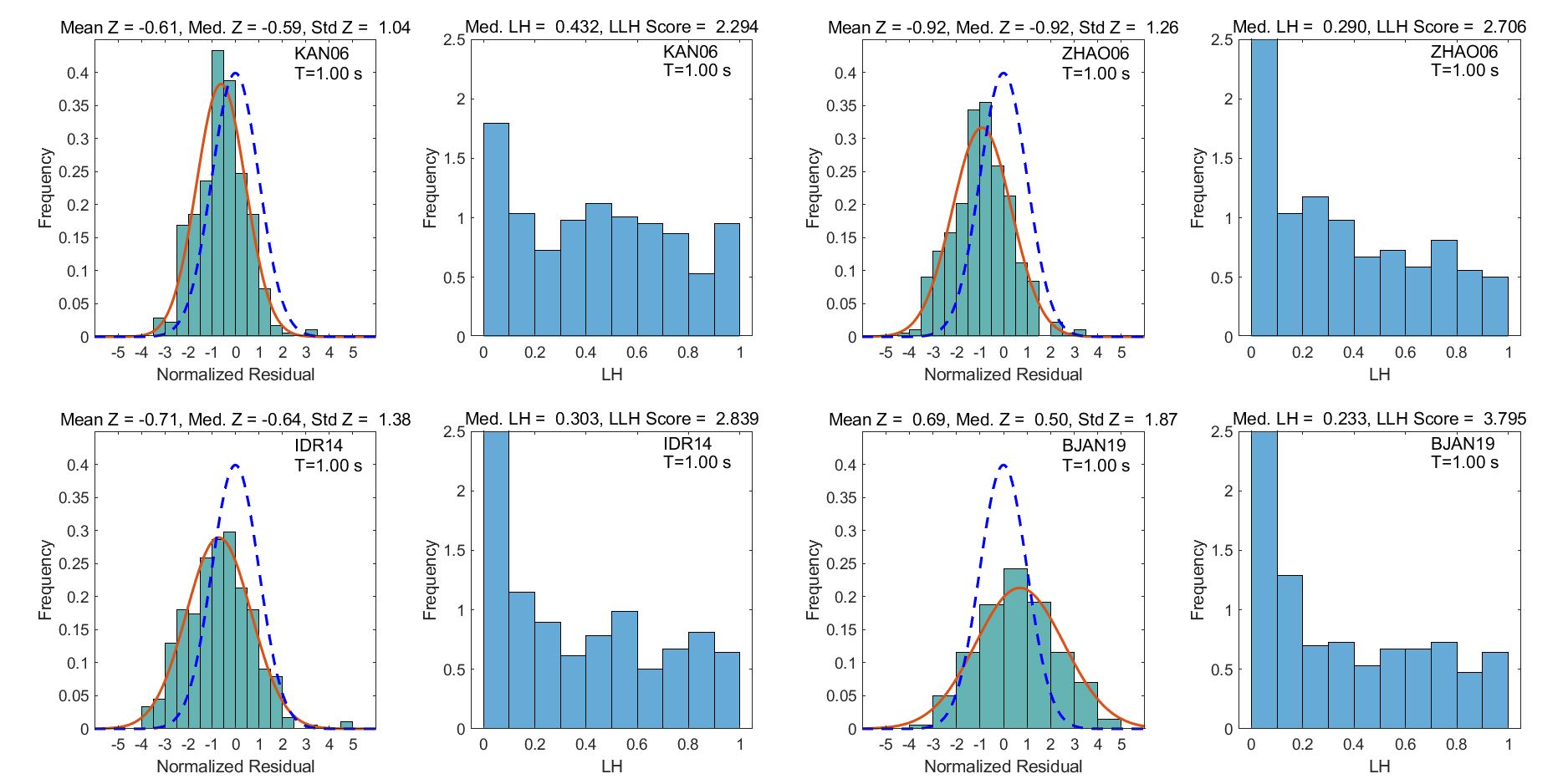}}} \\
    \end{tabular}
		\end{adjustbox}
\caption{Left panel of each subfigure: Graph showing the distribution of normalized residual and
the best fit normal distribution (continuous curve) with the standard
normal distribution (dashed curve) at three periods of interest for four best
	suitable GMPEs in western and central Himalayas. Right side of each 
	subfigure: Corresponding histograms 
	of the LH values with the median LH and LLH score.}
\label{lhwch}
\end{figure}

Table \ref{rewch} presents the various goodness-of-fit measures at 9
different periods for all the GMPEs considered. The values for the period $0.01$ s
may be treated as the values for PGA. PSA at $0.2$ s and $1.0$ s 
are frequently used as corner spectral periods to
construct design spectrum for structural design.
{\fontsize{8}{9}\selectfont
	\begin{longtable}{llccccccccc}
        \caption{Various goodness-of-fit measures for all the GMPEs considered
		in western and central Himalayas} \\ 
\hline\noalign{\smallskip}
	GMPEs      & Goodness-of-fit & & & & & Periods (s) & & & & \\
	Considered & measures & 0.01 & 0.02 & 0.05 & 0.10 & 0.20 & 0.30 & 0.50 & 1.00 & 2.00  \\
\noalign{\smallskip}\hline\noalign{\smallskip}
 KAN06 & Mean $Z$   &   0.25 &   0.29 &   0.29 &   0.29 &   0.18 &  -0.10 &  -0.40 &  -0.61 &  -0.48 \\ 
      &  Median $Z$ &   0.26 &   0.29 &   0.28 &   0.23 &   0.20 &  -0.08 &  -0.47 &  -0.59 &  -0.64 \\ 
       & Std $Z$    &   1.09 &   1.09 &   1.16 &   1.08 &   1.16 &   1.17 &   1.07 &   1.04 &   1.24 \\ 
       & Med. $LH$  &   0.47 &   0.48 &   0.44 &   0.45 &   0.40 &   0.38 &   0.42 &   0.43 &   0.34 \\ 
       & Rank        & B     & B     & B     & B     & B     & B     & B     & C     & C    \\ 
       & $LLH$ Score&   2.00 &   2.02 &   2.13 &   2.11 &   2.21 &   2.17 &   2.18 &   2.29 &   2.44 \\ 
\noalign{\smallskip}\hline\noalign{\smallskip}
ZHAO06 & Mean $Z$   &  -0.08 &  -0.23 &  -0.06 &   0.03 &  -0.01 &  -0.30 &  -0.75 &  -0.92 &  -0.86 \\ 
      &  Median $Z$ &  -0.07 &  -0.21 &  -0.05 &   0.01 &   0.03 &  -0.29 &  -0.79 &  -0.92 &  -0.98 \\ 
       & Std $Z$    &   1.23 &   1.21 &   1.22 &   1.10 &   1.27 &   1.34 &   1.30 &   1.26 &   1.37 \\ 
       & Med. $LH$  &   0.41 &   0.42 &   0.47 &   0.44 &   0.42 &   0.31 &   0.29 &   0.29 &   0.22 \\ 
       & Rank        & B     & B     & B     & A     & C     & C     & D     & D     & D    \\ 
       & $LLH$ Score&   1.95 &   2.00 &   2.03 &   1.97 &   2.18 &   2.31 &   2.56 &   2.71 &   2.88 \\ 
\noalign{\smallskip}\hline\noalign{\smallskip}
  AS08 & Mean $Z$   &  -0.29 &  -0.26 &  -0.08 &   0.05 &  -0.25 &  -0.58 &  -0.93 &  -1.01 &  -0.68 \\ 
      &  Median $Z$ &  -0.31 &  -0.29 &  -0.14 &  -0.04 &  -0.21 &  -0.56 &  -0.92 &  -0.98 &  -0.84 \\ 
       & Std $Z$    &   1.48 &   1.48 &   1.42 &   1.44 &   1.54 &   1.47 &   1.39 &   1.50 &   1.82 \\ 
       & Med. $LH$  &   0.33 &   0.32 &   0.41 &   0.33 &   0.32 &   0.26 &   0.24 &   0.23 &   0.20 \\ 
       & Rank        & C     & C     & C     & C     & D     & C     & D     & D     & D    \\ 
       & $LLH$ Score&   2.45 &   2.43 &   2.41 &   2.49 &   2.73 &   2.73 &   2.91 &   3.19 &   3.52 \\ 
\noalign{\smallskip}\hline\noalign{\smallskip}
  BA08 & Mean $Z$   &   0.16 &   0.16 &   0.40 &   0.37 &  -0.03 &  -0.48 &  -1.07 &  -1.02 &  -0.24 \\ 
      &  Median $Z$ &   0.15 &   0.12 &   0.33 &   0.39 &   0.06 &  -0.45 &  -1.03 &  -1.05 &  -0.81 \\ 
       & Std $Z$    &   1.75 &   1.75 &   1.79 &   1.65 &   1.65 &   1.57 &   1.57 &   1.64 &   2.23 \\ 
       & Med. $LH$  &   0.27 &   0.29 &   0.31 &   0.27 &   0.31 &   0.25 &   0.19 &   0.17 &   0.13 \\ 
       & Rank        & D     & D     & D     & D     & D     & D     & D     & D     & D    \\ 
       & $LLH$ Score&   2.73 &   2.72 &   2.97 &   2.67 &   2.54 &   2.55 &   3.22 &   3.40 &   4.42 \\ 
\noalign{\smallskip}\hline\noalign{\smallskip}
  CB08 & Mean $Z$   &  -0.78 &  -0.75 &  -0.51 &  -0.25 &  -0.68 &  -1.08 &  -1.51 &  -1.07 &  -0.06 \\ 
      &  Median $Z$ &  -0.75 &  -0.74 &  -0.49 &  -0.26 &  -0.58 &  -1.05 &  -1.51 &  -1.07 &  -0.50 \\ 
       & Std $Z$    &   1.66 &   1.66 &   1.63 &   1.51 &   1.67 &   1.73 &   1.70 &   1.67 &   2.16 \\ 
       & Med. $LH$  &   0.23 &   0.24 &   0.31 &   0.30 &   0.29 &   0.17 &   0.09 &   0.18 &   0.18 \\ 
       & Rank        & D     & D     & D     & D     & D     & D     & D     & D     & D    \\ 
       & $LLH$ Score&   2.83 &   2.78 &   2.62 &   2.29 &   2.91 &   3.55 &   4.28 &   3.48 &   4.04 \\ 
\noalign{\smallskip}\hline\noalign{\smallskip}
  CY08 & Mean $Z$   &   0.55 &   0.56 &   0.64 &   0.93 &   0.75 &   0.47 &   0.07 &  -0.28 &   0.14 \\ 
      &  Median $Z$ &   0.57 &   0.56 &   0.60 &   0.81 &   0.82 &   0.58 &   0.07 &  -0.43 &  -0.55 \\ 
       & Std $Z$    &   1.70 &   1.71 &   1.74 &   1.72 &   1.68 &   1.58 &   1.57 &   1.78 &   2.45 \\ 
       & Med. $LH$  &   0.28 &   0.26 &   0.30 &   0.21 &   0.24 &   0.28 &   0.31 &   0.21 &   0.16 \\ 
       & Rank        & D     & D     & D     & D     & D     & D     & D     & D     & D    \\ 
       & $LLH$ Score&   3.05 &   3.08 &   3.30 &   3.60 &   3.28 &   2.81 &   2.59 &   3.14 &   5.17 \\ 
\noalign{\smallskip}\hline\noalign{\smallskip}
 IDR08 & Mean $Z$   &  -0.22 &  -0.18 &  -0.04 &   0.36 &   0.01 &  -0.44 &  -0.91 &  -1.18 &  -0.86 \\ 
      &  Median $Z$ &  -0.14 &  -0.11 &  -0.03 &   0.36 &   0.13 &  -0.34 &  -0.95 &  -1.21 &  -1.02 \\ 
       & Std $Z$    &   1.41 &   1.41 &   1.40 &   1.30 &   1.41 &   1.39 &   1.30 &   1.21 &   1.25 \\ 
       & Med. $LH$  &   0.34 &   0.33 &   0.43 &   0.36 &   0.34 &   0.32 &   0.26 &   0.21 &   0.23 \\ 
       & Rank        & C     & C     & C     & C     & C     & C     & D     & D     & D    \\ 
       & $LLH$ Score&   2.30 &   2.27 &   2.23 &   2.21 &   2.39 &   2.52 &   2.86 &   3.16 &   2.82 \\ 
\noalign{\smallskip}\hline\noalign{\smallskip}
AKBO10 & Mean $Z$   &   0.38 &   0.28 &   0.59 &   0.51 &   0.01 &  -0.18 &  -0.23 &   0.16 &   0.43 \\ 
      &  Median $Z$ &   0.42 &   0.33 &   0.60 &   0.54 &   0.08 &  -0.10 &  -0.26 &  -0.10 &  -0.11 \\ 
       & Std $Z$    &   1.54 &   1.56 &   1.64 &   1.51 &   1.44 &   1.40 &   1.38 &   2.03 &   2.46 \\ 
       & Med. $LH$  &   0.32 &   0.36 &   0.31 &   0.32 &   0.38 &   0.33 &   0.38 &   0.22 &   0.17 \\ 
       & Rank        & D     & D     & D     & D     & C     & C     & C     & D     & D    \\ 
       & $LLH$ Score&   2.51 &   2.53 &   2.95 &   2.60 &   2.30 &   2.25 &   2.33 &   3.89 &   5.41 \\ 
\noalign{\smallskip}\hline\noalign{\smallskip}
 ASK14 & Mean $Z$   &   0.88 &   0.82 &   0.97 &   1.13 &   1.05 &   0.88 &   0.58 &   0.68 &   0.84 \\ 
      &  Median $Z$ &   0.90 &   0.86 &   0.99 &   1.14 &   1.13 &   0.91 &   0.52 &   0.53 &   0.33 \\ 
       & Std $Z$    &   1.32 &   1.32 &   1.34 &   1.35 &   1.40 &   1.33 &   1.32 &   1.76 &   2.32 \\ 
       & Med. $LH$  &   0.28 &   0.30 &   0.25 &   0.20 &   0.21 &   0.25 &   0.31 &   0.26 &   0.25 \\ 
       & Rank        & D     & D     & D     & D     & D     & D     & C     & D     & D    \\ 
       & $LLH$ Score&   2.81 &   2.76 &   3.03 &   3.30 &   3.26 &   2.87 &   2.49 &   3.51 &   5.30 \\ 
\noalign{\smallskip}\hline\noalign{\smallskip}
BSSA14 & Mean $Z$   &   1.14 &   1.15 &   1.14 &   1.22 &   1.19 &   1.00 &   0.68 &   0.65 &   1.08 \\ 
      &  Median $Z$ &   1.20 &   1.22 &   1.21 &   1.24 &   1.32 &   1.03 &   0.64 &   0.49 &   0.38 \\ 
       & Std $Z$    &   1.47 &   1.45 &   1.41 &   1.44 &   1.45 &   1.43 &   1.48 &   1.86 &   2.43 \\ 
       & Med. $LH$  &   0.16 &   0.16 &   0.19 &   0.17 &   0.14 &   0.19 &   0.30 &   0.29 &   0.32 \\ 
       & Rank        & D     & D     & D     & D     & D     & D     & C     & D     & D    \\ 
       & $LLH$ Score&   3.37 &   3.37 &   3.43 &   3.57 &   3.44 &   3.06 &   2.76 &   3.66 &   5.98 \\ 
\noalign{\smallskip}\hline\noalign{\smallskip}
  CB14 & Mean $Z$   &   1.49 &   1.48 &   1.29 &   1.54 &   1.50 &   1.07 &   0.65 &   0.43 &   0.89 \\ 
      &  Median $Z$ &   1.49 &   1.49 &   1.37 &   1.62 &   1.61 &   1.07 &   0.62 &   0.29 &   0.17 \\ 
       & Std $Z$    &   1.60 &   1.59 &   1.45 &   1.44 &   1.49 &   1.44 &   1.48 &   1.91 &   2.55 \\ 
       & Med. $LH$  &   0.12 &   0.12 &   0.14 &   0.09 &   0.10 &   0.20 &   0.30 &   0.25 &   0.29 \\ 
       & Rank        & D     & D     & D     & D     & D     & D     & C     & D     & D    \\ 
       & $LLH$ Score&   4.32 &   4.30 &   3.74 &   4.23 &   4.15 &   3.22 &   2.73 &   3.64 &   6.09 \\ 
\noalign{\smallskip}\hline\noalign{\smallskip}
  CY14 & Mean $Z$   &   1.19 &   1.21 &   1.11 &   1.31 &   1.22 &   0.92 &   0.65 &   0.58 &   0.93 \\ 
      &  Median $Z$ &   1.25 &   1.27 &   1.11 &   1.34 &   1.24 &   0.89 &   0.58 &   0.42 &   0.24 \\ 
       & Std $Z$    &   1.47 &   1.48 &   1.56 &   1.52 &   1.45 &   1.35 &   1.33 &   1.80 &   2.54 \\ 
       & Med. $LH$  &   0.17 &   0.17 &   0.18 &   0.14 &   0.15 &   0.23 &   0.32 &   0.30 &   0.30 \\ 
       & Rank        & D     & D     & D     & D     & D     & D     & C     & D     & D    \\ 
       & $LLH$ Score&   3.46 &   3.50 &   3.57 &   3.86 &   3.59 &   2.94 &   2.62 &   3.57 &   6.17 \\ 
\noalign{\smallskip}\hline\noalign{\smallskip}
 IDR14 & Mean $Z$   &  -0.02 &  -0.02 &   0.14 &   0.34 &   0.20 &  -0.08 &  -0.37 &  -0.70 &  -0.41 \\ 
      &  Median $Z$ &   0.00 &  -0.01 &   0.07 &   0.29 &   0.31 &  -0.09 &  -0.34 &  -0.64 &  -0.37 \\ 
       & Std $Z$    &   1.60 &   1.59 &   1.61 &   1.49 &   1.56 &   1.54 &   1.41 &   1.38 &   1.44 \\ 
       & Med. $LH$  &   0.24 &   0.24 &   0.28 &   0.28 &   0.24 &   0.24 &   0.33 &   0.30 &   0.40 \\ 
       & Rank        & D     & D     & D     & C     & D     & D     & C     & C     & C    \\ 
       & $LLH$ Score&   2.76 &   2.74 &   2.81 &   2.65 &   2.79 &   2.75 &   2.60 &   2.84 &   2.75 \\ 
\noalign{\smallskip}\hline\noalign{\smallskip}
  GK15 & Mean $Z$   &   0.26 &   0.30 &   0.15 &   0.35 &   0.31 &   0.00 &  -0.62 &  -1.10 &  -0.92 \\ 
      &  Median $Z$ &   0.10 &   0.13 &  -0.08 &   0.13 &   0.24 &  -0.04 &  -0.65 &  -1.23 &  -1.15 \\ 
       & Std $Z$    &   1.69 &   1.73 &   1.84 &   1.76 &   1.68 &   1.71 &   1.71 &   1.62 &   1.71 \\ 
       & Med. $LH$  &   0.33 &   0.33 &   0.32 &   0.34 &   0.34 &   0.24 &   0.17 &   0.15 &   0.17 \\ 
       & Rank        & D     & D     & D     & D     & D     & D     & D     & D     & D    \\ 
       & $LLH$ Score&   2.83 &   2.92 &   3.17 &   3.04 &   2.82 &   2.85 &   3.19 &   3.76 &   3.86 \\ 
\noalign{\smallskip}\hline\noalign{\smallskip}
ZHAO16 & Mean $Z$   &  -0.17 &  -0.17 &  -0.41 &  -0.27 &   0.23 &   0.32 &   0.25 &   0.09 &   0.01 \\ 
      &  Median $Z$ &  -0.15 &  -0.16 &  -0.46 &  -0.32 &   0.28 &   0.38 &   0.28 &   0.14 &  -0.17 \\ 
       & Std $Z$    &   1.38 &   1.38 &   1.37 &   1.21 &   1.27 &   1.26 &   1.25 &   1.29 &   1.53 \\ 
       & Med. $LH$  &   0.37 &   0.37 &   0.41 &   0.42 &   0.38 &   0.39 &   0.40 &   0.38 &   0.31 \\ 
       & Rank        & C     & C     & C     & B     & C     & C     & B     & C     & D    \\ 
       & $LLH$ Score&   2.15 &   2.15 &   2.33 &   2.09 &   2.17 &   2.17 &   2.09 &   2.15 &   2.57 \\ 
\noalign{\smallskip}\hline\noalign{\smallskip}
BJAN19 & Mean $Z$   &   0.41 &   0.26 &   0.02 &   0.23 &   0.36 &   0.28 &   0.42 &   0.69 &   1.39 \\ 
      &  Median $Z$ &   0.39 &   0.22 &  -0.12 &   0.07 &   0.33 &   0.31 &   0.42 &   0.51 &   0.89 \\ 
       & Std $Z$    &   1.16 &   1.12 &   1.15 &   1.11 &   1.18 &   1.25 &   1.39 &   1.87 &   2.50 \\ 
       & Med. $LH$  &   0.44 &   0.49 &   0.48 &   0.48 &   0.43 &   0.39 &   0.32 &   0.23 &   0.19 \\ 
       & Rank        & B     & B     & B     & A     & B     & B     & C     & D     & D    \\ 
       & $LLH$ Score&   2.12 &   2.08 &   2.17 &   2.14 &   2.15 &   2.17 &   2.46 &   3.79 &   6.79 \\ 
\noalign{\smallskip}\hline
\label{rewch} \\
\end{longtable}
}
It is evident from Table \ref{rewch} that each of the goodness-of-fit 
measures varies considerably with the periods for all the GMPEs and
therefore the performance of GMPEs are period-dependent.
It is also
observed that the NGA-West1 GMPEs, except AS08 and IDR08, fall into 
`unacceptable category' for all the $9$ periods in this target region. The
AS08 and the IDR08 show `low capability' at lower periods. The NGA-West2
GMPEs are not suitable too for most of the periods. However, ASK14 and IDR14
show better LLH scores than the other NGA-West2 GMPEs. It is interesting to
note that the NGA-West1 GMPEs perform better than the NGA-West2 GMPEs in terms
of LLH score, especially at lower periods, given the set of available data in
this target region. 
As per the ranking criteria in LH method, KAN06 
shows no `unacceptable category' at any period. 
From our analysis, a combination of 
ZHAO06, KAN06, BJAN19 and ZHAO16 are seen to best perform at $0.01$ s and may
be selected as a suitable combination for this period, compared to the other GMPEs, 
for populating the branches of the logic tree in PSHA.
Since LLH method is considered
better than LH method (as explained earlier), ZHAO06 may be treated as the most
suitable GMPE for periods $\leq$ $0.1$ s and above $0.1$ s, ZHAO16 may be 
considered as the most suitable in this target region, although at $2$ s KAN06
shows lowest LLH score. For $0.2$ s, a combination of BJAN19, ZHAO16, ZHAO06
and KAN06 may be considered appropriate for designing the logic tree 
in PSHA for this target region. And for $1.0$ s, a combination of ZHAO16, 
KAN06, ZHAO06 and IDR14 may be considered suitable for construction of logic
tree in carrying out PSHA.
The variation of
LLH scores against all the periods for all the GMPES are plotted 
in Fig. \ref{llhper1}.
For clarity and easy visualization, the bar diagrams of LLH scores at three
periods of importance for PSHA for all the GMPEs considered are presented
in Fig. \ref{llhper2}.
The distribution of normalized residuals, 
the best fit normal distribution (continuous curve) with the standard
normal distribution (dashed curve) and the corresponding histograms of
the LH values
are presented in Fig. \ref{lhwch} for the suitable GMPEs at periods $0.01$ s, 
$0.2$ s and $1.0$ s respectively. It may be noted that the median $LH$ values
and the $LLH$ scores are consistent for all the suitable set of GMPEs. 

\subsection{Results for Indo-Gangetic Plains}
\begin{figure}[!h]
\begin{center}
        \rotatebox{0}{\includegraphics[scale=0.6]{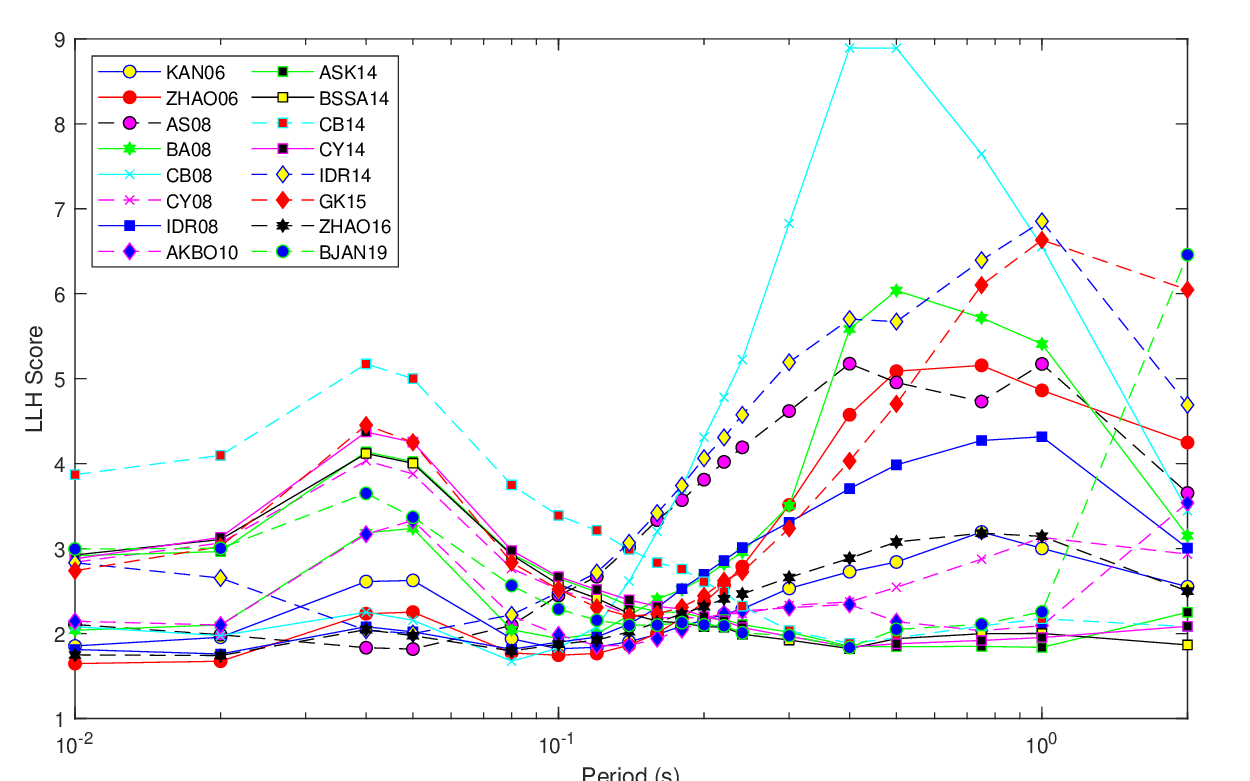}}
	\caption{Plot of LLH score at 19 periods 
	for the all the GMPEs considered in Indo-Gangetic Plains.}
\label{llhper3}
\end{center}
\end{figure}

\begin{figure}[!h]
\begin{center}
        \rotatebox{0}{\includegraphics[scale=0.6]{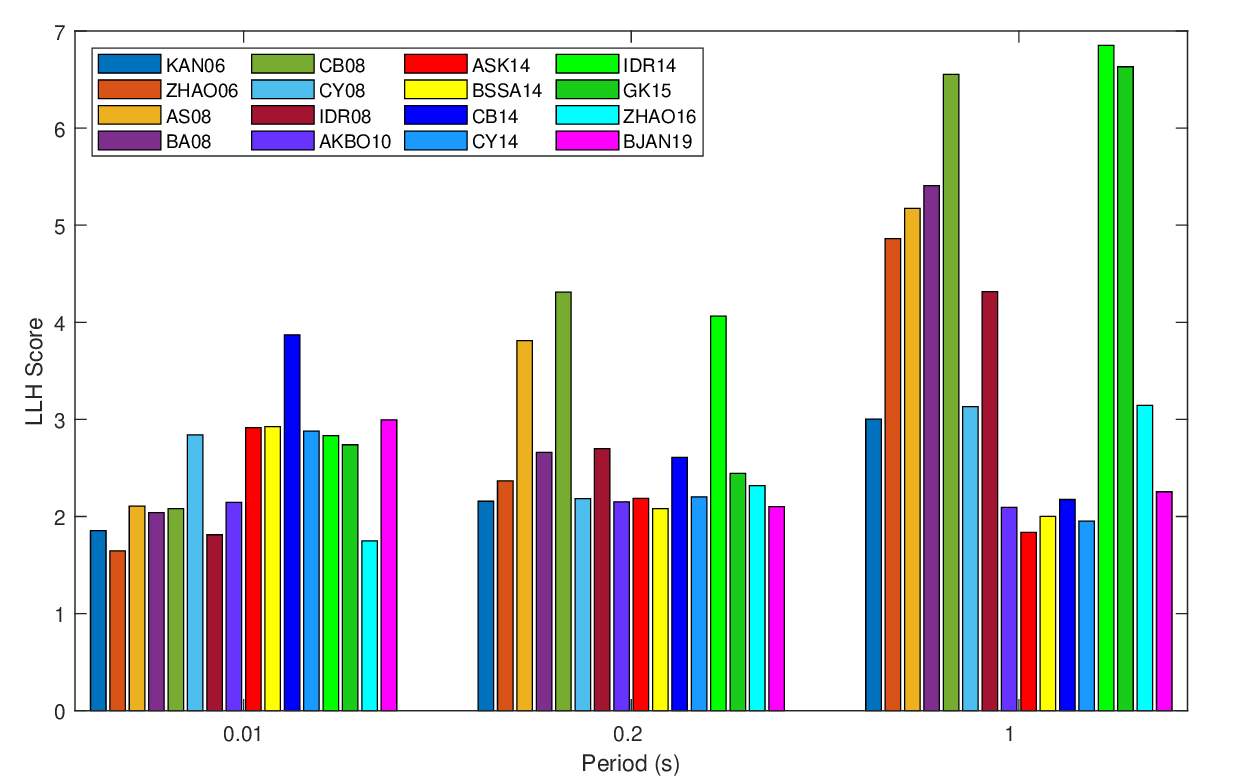}}
	\caption{Bar diagram of LLH score at $0.01$s, $0.2$s and 
	$1.0$s for the all the GMPEs considered in Indo-Gangetic
	Plains.}
\label{llhper4}
\end{center}
\end{figure}

\begin{figure}[!h]
        \begin{adjustbox}{max width=1.1\textwidth,center}
\begin{tabular}{c}
      \resizebox{120mm}{!}{\rotatebox{0}{\includegraphics[scale=0.6]{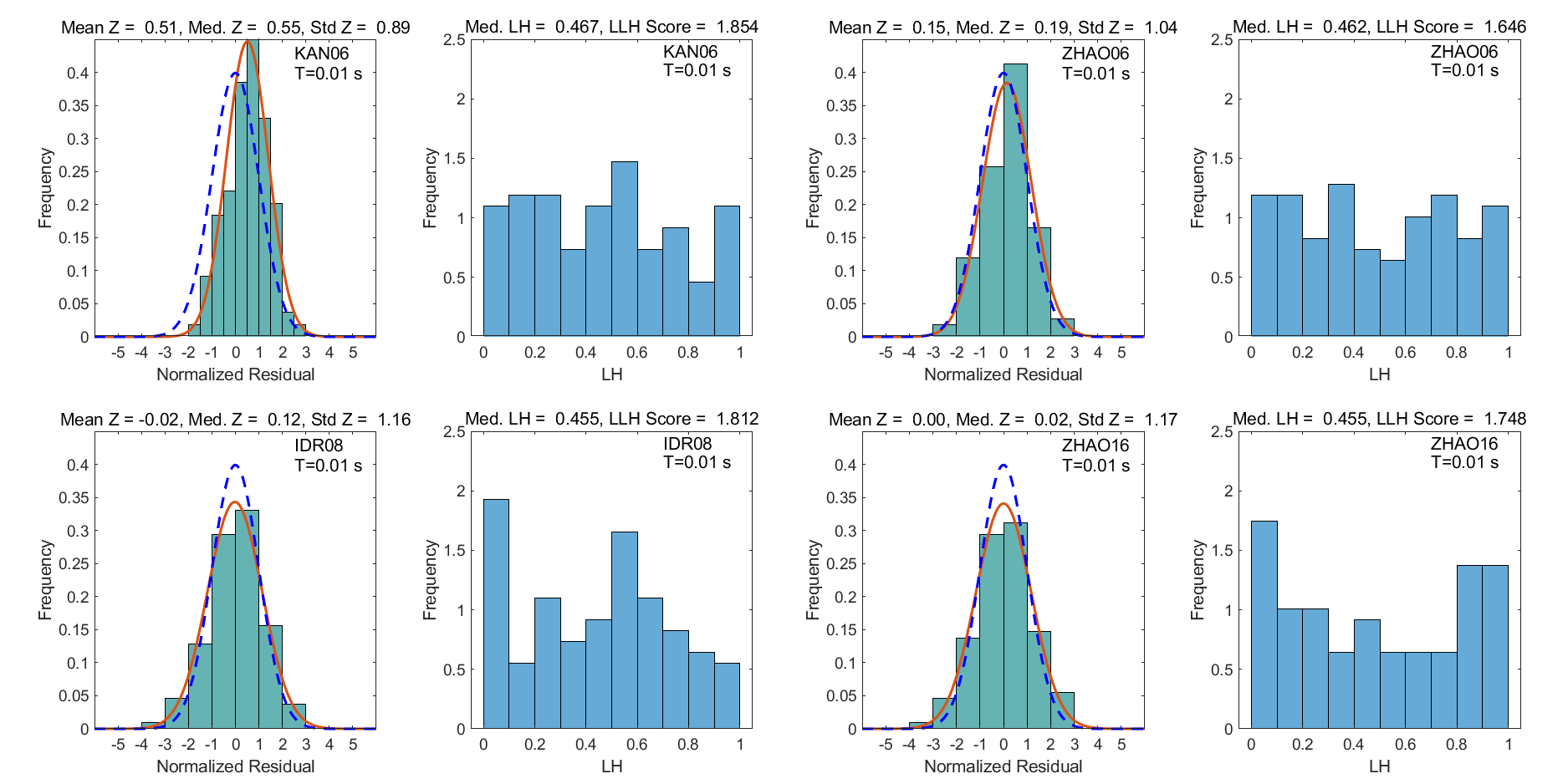}}} \\
      \resizebox{120mm}{!}{\rotatebox{0}{\includegraphics[scale=0.6]{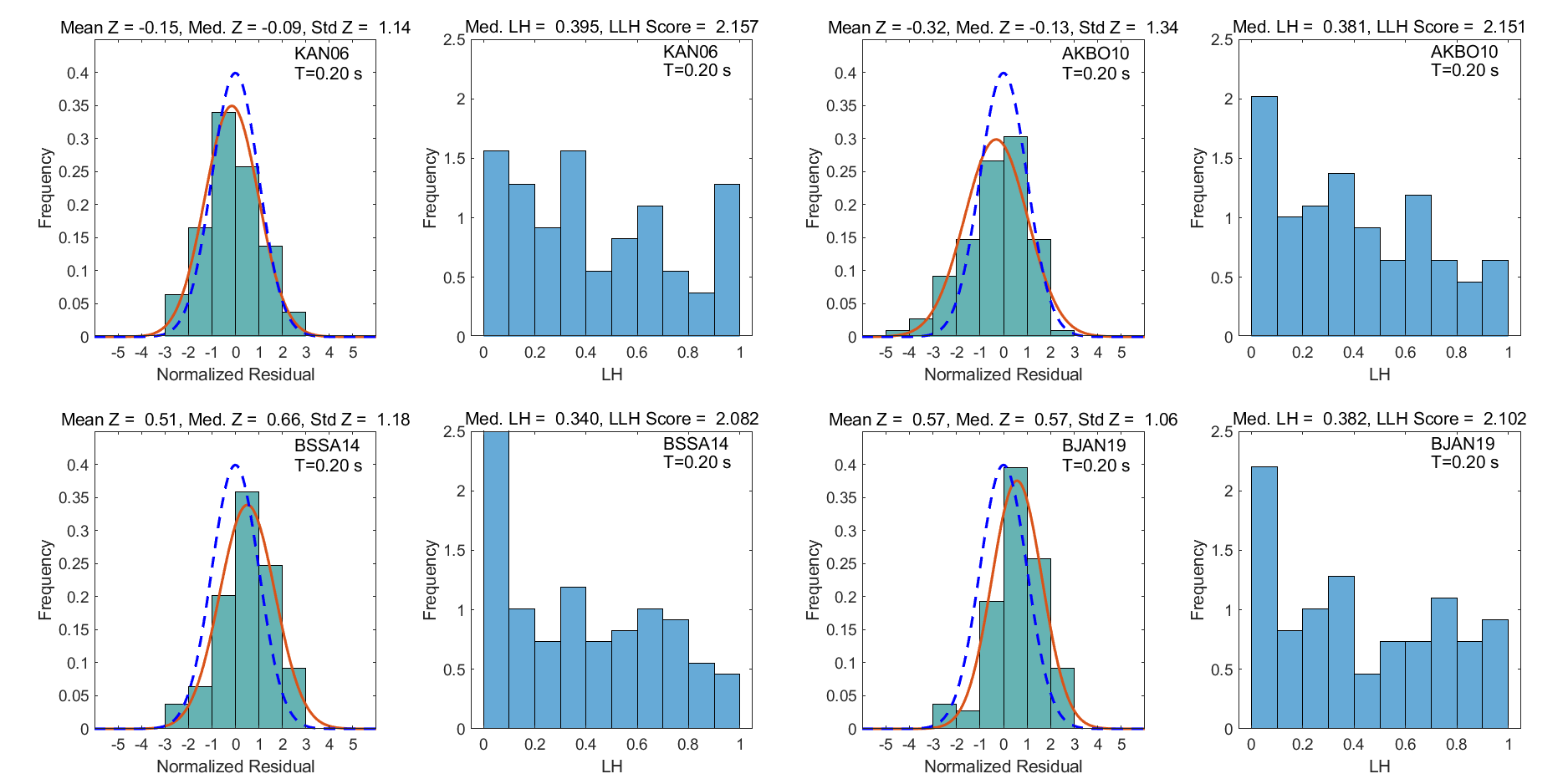}}} \\
      \resizebox{120mm}{!}{\rotatebox{0}{\includegraphics[scale=0.6]{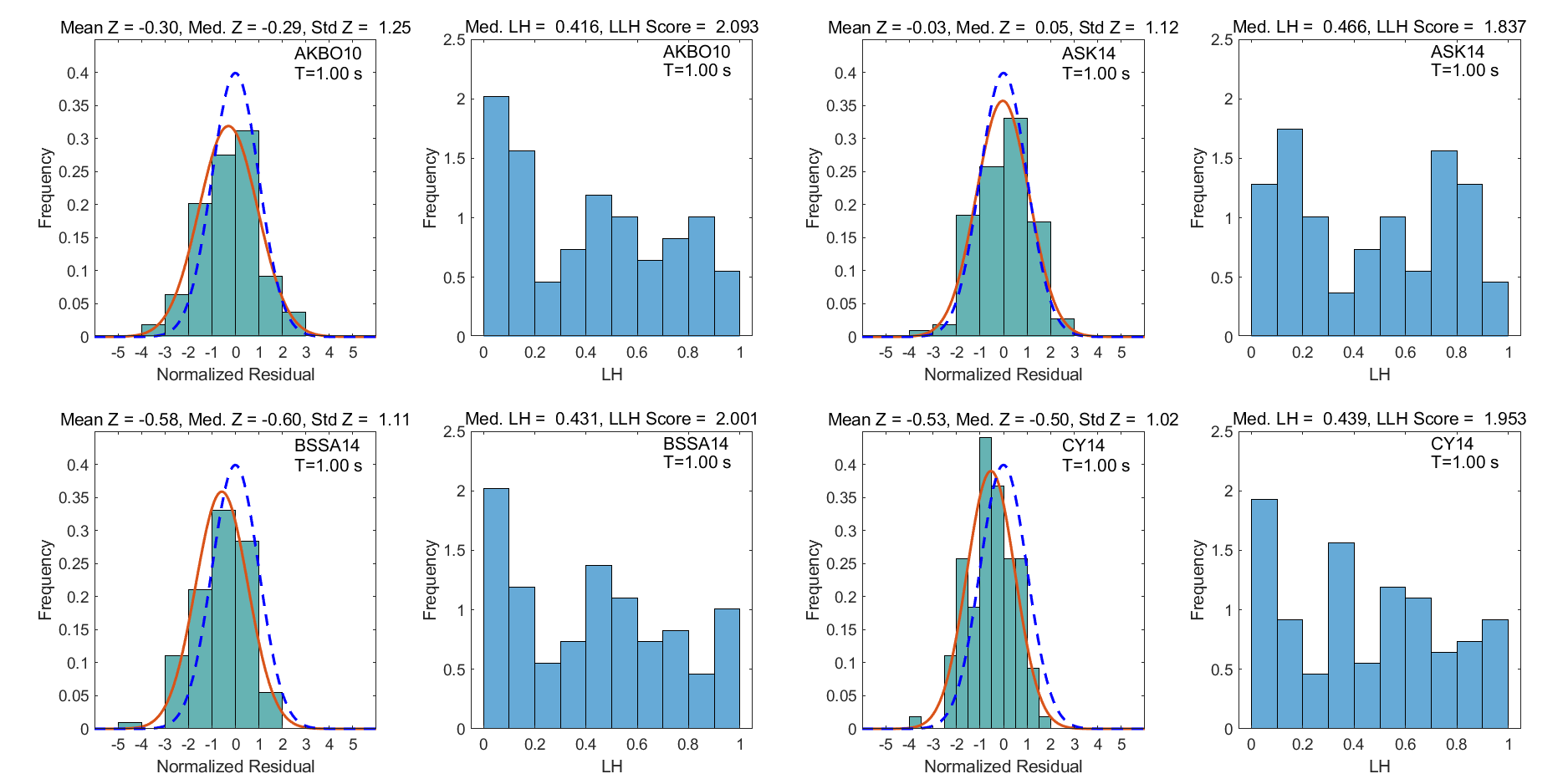}}} \\
    \end{tabular}
                \end{adjustbox}
\caption{Left panel of each subfigure: Graph showing the distribution of 
	normalized residual and
the best fit normal distribution (continuous curve) with the standard
normal distribution (dashed curve) at three periods of interest for four best
        suitable GMPEs in Indo-Gangetic Plains. Right side of each
        subfigure: Corresponding histograms
        of the LH values with the median LH and LLH score.}
\label{lhwch1}
\end{figure}

Table \ref{reIG} presents the overall summary of the goodness-of-fit
parameters at $9$ different periods for all the GMPEs considered.
{\fontsize{8}{9}\selectfont
	\begin{longtable}{llccccccccc}
        \caption{Various goodness-of-fit measures
        for all the GMPEs considered in Indo-Gangetic Plains} \\ 
\hline\noalign{\smallskip}
	GMPEs      & Goodness-of-fit & & & & & Periods (s) & & & & \\
	Considered & measures & 0.01 & 0.02 & 0.05 & 0.10 & 0.20 & 0.30 & 0.50 & 1.00 & 2.00  \\
\noalign{\smallskip}\hline\noalign{\smallskip}
KAN06 & Mean $Z$   &   0.51 &   0.62 &   1.02 &   0.33 &  -0.15 &  -0.58 &  -0.88 &  -1.15 &  -1.10 \\
      &  Median $Z$ &   0.55 &   0.69 &   1.05 &   0.32 &  -0.09 &  -0.74 &  -0.90 &  -1.27 &  -1.22 \\
       & Std $Z$    &   0.89 &   0.91 &   1.04 &   0.86 &   1.14 &   1.25 &   1.21 &   1.06 &   0.84 \\
       & Med. $LH$  &   0.47 &   0.43 &   0.29 &   0.54 &   0.40 &   0.33 &   0.22 &   0.20 &   0.22 \\
       & Rank        & C     & C     & D     & B     & B     & C     & D     & D     & D    \\
       & $LLH$ Score&   1.85 &   1.96 &   2.63 &   1.82 &   2.16 &   2.53 &   2.84 &   3.00 &   2.55 \\
\noalign{\smallskip}\hline\noalign{\smallskip}
ZHAO06 & Mean $Z$   &   0.15 &   0.13 &   0.73 &   0.18 &  -0.45 &  -1.19 &  -1.85 &  -1.85 &  -1.83 \\
      &  Median $Z$ &   0.20 &   0.15 &   0.75 &   0.22 &  -0.45 &  -1.45 &  -1.93 &  -2.13 &  -2.04 \\
       & Std $Z$    &   1.04 &   1.03 &   1.12 &   0.94 &   1.29 &   1.47 &   1.54 &   1.41 &   1.10 \\
       & Med. $LH$  &   0.46 &   0.51 &   0.34 &   0.55 &   0.31 &   0.13 &   0.05 &   0.03 &   0.04 \\
       & Rank        & A     & A     & D     & A     & C     & D     & D     & D     & D    \\
       & $LLH$ Score&   1.65 &   1.67 &   2.25 &   1.75 &   2.37 &   3.51 &   5.09 &   4.86 &   4.25 \\
\noalign{\smallskip}\hline\noalign{\smallskip}
  AS08 & Mean $Z$   &  -0.65 &  -0.54 &  -0.01 &  -0.60 &  -1.20 &  -1.63 &  -1.89 &  -2.08 &  -1.61 \\
      &  Median $Z$ &  -0.65 &  -0.52 &  -0.01 &  -0.55 &  -1.23 &  -1.84 &  -1.91 &  -1.96 &  -1.43 \\
       & Std $Z$    &   1.18 &   1.17 &   1.10 &   1.29 &   1.59 &   1.57 &   1.44 &   1.30 &   1.17 \\
       & Med. $LH$  &   0.32 &   0.39 &   0.45 &   0.33 &   0.16 &   0.06 &   0.06 &   0.05 &   0.15 \\
       & Rank        & C     & C     & A     & C     & D     & D     & D     & D     & D    \\
       & $LLH$ Score&   2.11 &   1.99 &   1.82 &   2.45 &   3.81 &   4.62 &   4.96 &   5.17 &   3.65 \\
\noalign{\smallskip}\hline\noalign{\smallskip}
  BA08 & Mean $Z$   &   0.01 &   0.16 &   0.97 &   0.06 &  -0.71 &  -1.34 &  -2.19 &  -2.20 &  -1.34 \\
      &  Median $Z$ &   0.05 &   0.16 &   0.87 &  -0.06 &  -0.70 &  -1.24 &  -2.21 &  -2.18 &  -1.37 \\
       & Std $Z$    &   1.47 &   1.49 &   1.67 &   1.36 &   1.55 &   1.50 &   1.66 &   1.30 &   1.20 \\
       & Med. $LH$  &   0.29 &   0.32 &   0.18 &   0.37 &   0.28 &   0.15 &   0.03 &   0.03 &   0.16 \\
       & Rank        & C     & C     & D     & C     & D     & D     & D     & D     & D    \\
       & $LLH$ Score&   2.04 &   2.10 &   3.24 &   1.94 &   2.66 &   3.50 &   6.04 &   5.41 &   3.15 \\
\noalign{\smallskip}\hline\noalign{\smallskip}
  CB08 & Mean $Z$   &  -0.73 &  -0.57 &   0.18 &  -0.54 &  -1.50 &  -2.25 &  -2.78 &  -2.45 &  -1.60 \\
      &  Median $Z$ &  -0.65 &  -0.50 &   0.07 &  -0.51 &  -1.34 &  -2.38 &  -2.67 &  -2.53 &  -1.60 \\
       & Std $Z$    &   1.36 &   1.37 &   1.51 &   1.21 &   1.73 &   1.92 &   1.97 &   1.49 &   1.13 \\
       & Med. $LH$  &   0.33 &   0.36 &   0.33 &   0.36 &   0.14 &   0.02 &   0.01 &   0.01 &   0.11 \\
       & Rank        & C     & C     & D     & C     & D     & D     & D     & D     & D    \\
       & $LLH$ Score&   2.08 &   1.98 &   2.16 &   1.83 &   4.31 &   6.83 &   8.89 &   6.55 &   3.44 \\
\noalign{\smallskip}\hline\noalign{\smallskip}
  CY08 & Mean $Z$   &   0.59 &   0.69 &   1.15 &   0.61 &   0.06 &  -0.38 &  -0.80 &  -1.30 &  -1.12 \\
      &  Median $Z$ &   0.59 &   0.67 &   1.14 &   0.47 &   0.14 &  -0.39 &  -0.74 &  -1.17 &  -1.19 \\
       & Std $Z$    &   1.60 &   1.66 &   1.71 &   1.40 &   1.36 &   1.39 &   1.33 &   1.24 &   1.30 \\
       & Med. $LH$  &   0.23 &   0.20 &   0.15 &   0.31 &   0.38 &   0.37 &   0.29 &   0.23 &   0.20 \\
       & Rank        & D     & D     & D     & C     & C     & C     & D     & D     & D    \\
       & $LLH$ Score&   2.84 &   3.07 &   3.88 &   2.52 &   2.18 &   2.33 &   2.54 &   3.13 &   2.94 \\
\noalign{\smallskip}\hline\noalign{\smallskip}
 IDR08 & Mean $Z$   &  -0.02 &   0.10 &   0.63 &   0.30 &  -0.31 &  -0.86 &  -1.29 &  -1.61 &  -1.32 \\
      &  Median $Z$ &   0.12 &   0.18 &   0.64 &   0.51 &  -0.21 &  -0.89 &  -1.18 &  -1.62 &  -1.31 \\
       & Std $Z$    &   1.16 &   1.13 &   1.11 &   1.14 &   1.52 &   1.57 &   1.56 &   1.36 &   0.89 \\
       & Med. $LH$  &   0.46 &   0.49 &   0.38 &   0.40 &   0.26 &   0.23 &   0.14 &   0.11 &   0.19 \\
       & Rank        & B     & B     & C     & C     & D     & D     & D     & D     & D    \\
       & $LLH$ Score&   1.81 &   1.76 &   2.02 &   1.90 &   2.70 &   3.31 &   3.98 &   4.32 &   3.00 \\
\noalign{\smallskip}\hline\noalign{\smallskip}
AKBO10 & Mean $Z$   &   0.60 &   0.52 &   1.21 &   0.47 &  -0.32 &  -0.62 &  -0.63 &  -0.30 &  -0.01 \\
      &  Median $Z$ &   0.75 &   0.67 &   1.10 &   0.62 &  -0.14 &  -0.42 &  -0.52 &  -0.29 &  -0.41 \\
       & Std $Z$    &   1.29 &   1.29 &   1.45 &   1.21 &   1.34 &   1.30 &   1.14 &   1.25 &   1.91 \\
       & Med. $LH$  &   0.34 &   0.34 &   0.19 &   0.36 &   0.38 &   0.30 &   0.45 &   0.42 &   0.26 \\
       & Rank        & C     & C     & D     & C     & C     & C     & C     & C     & D    \\
       & $LLH$ Score&   2.15 &   2.10 &   3.33 &   1.98 &   2.15 &   2.31 &   2.14 &   2.09 &   3.54 \\
\noalign{\smallskip}\hline\noalign{\smallskip}
 ASK14 & Mean $Z$   &   1.08 &   1.08 &   1.54 &   1.01 &   0.43 &   0.05 &  -0.14 &  -0.03 &   0.14 \\
      &  Median $Z$ &   1.16 &   1.20 &   1.42 &   1.00 &   0.43 &   0.04 &  -0.04 &   0.05 &   0.05 \\
       & Std $Z$    &   1.21 &   1.23 &   1.31 &   1.07 &   1.16 &   1.16 &   1.08 &   1.12 &   1.37 \\
       & Med. $LH$  &   0.18 &   0.17 &   0.12 &   0.27 &   0.38 &   0.44 &   0.48 &   0.47 &   0.42 \\
       & Rank        & D     & D     & D     & D     & B     & B     & A     & A     & C    \\
       & $LLH$ Score&   2.92 &   2.96 &   4.02 &   2.65 &   2.19 &   2.02 &   1.84 &   1.84 &   2.25 \\
\noalign{\smallskip}\hline\noalign{\smallskip}
BSSA14 & Mean $Z$   &   1.05 &   1.14 &   1.52 &   0.96 &   0.51 &   0.06 &  -0.36 &  -0.58 &  -0.40 \\
      &  Median $Z$ &   1.07 &   1.21 &   1.53 &   0.92 &   0.66 &   0.14 &  -0.29 &  -0.60 &  -0.51 \\
       & Std $Z$    &   1.32 &   1.32 &   1.32 &   1.11 &   1.18 &   1.22 &   1.20 &   1.11 &   1.10 \\
       & Med. $LH$  &   0.18 &   0.18 &   0.12 &   0.27 &   0.34 &   0.36 &   0.45 &   0.43 &   0.37 \\
       & Rank        & D     & D     & D     & D     & C     & B     & B     & C     & C    \\
       & $LLH$ Score&   2.92 &   3.11 &   4.00 &   2.58 &   2.08 &   1.92 &   1.94 &   2.00 &   1.87 \\
\noalign{\smallskip}\hline\noalign{\smallskip}
  CB14 & Mean $Z$   &   1.61 &   1.69 &   1.97 &   1.46 &   0.91 &   0.21 &  -0.30 &  -0.73 &  -0.57 \\
      &  Median $Z$ &   1.83 &   1.89 &   2.04 &   1.52 &   0.97 &   0.26 &  -0.18 &  -0.65 &  -0.69 \\
       & Std $Z$    &   1.22 &   1.24 &   1.25 &   1.06 &   1.21 &   1.23 &   1.18 &   1.12 &   1.18 \\
       & Med. $LH$  &   0.07 &   0.06 &   0.04 &   0.12 &   0.25 &   0.35 &   0.45 &   0.39 &   0.34 \\
       & Rank        & D     & D     & D     & D     & D     & B     & B     & C     & C    \\
       & $LLH$ Score&   3.87 &   4.10 &   5.00 &   3.39 &   2.61 &   2.03 &   1.95 &   2.17 &   2.08 \\
\noalign{\smallskip}\hline\noalign{\smallskip}
  CY14 & Mean $Z$   &   1.13 &   1.24 &   1.63 &   1.07 &   0.59 &   0.08 &  -0.25 &  -0.53 &  -0.53 \\
      &  Median $Z$ &   1.06 &   1.24 &   1.61 &   0.94 &   0.64 &   0.14 &  -0.24 &  -0.50 &  -0.67 \\
       & Std $Z$    &   1.21 &   1.25 &   1.39 &   1.09 &   1.14 &   1.13 &   1.04 &   1.02 &   1.16 \\
       & Med. $LH$  &   0.19 &   0.14 &   0.10 &   0.23 &   0.38 &   0.39 &   0.48 &   0.44 &   0.34 \\
       & Rank        & D     & D     & D     & D     & C     & B     & B     & C     & C    \\
       & $LLH$ Score&   2.88 &   3.13 &   4.25 &   2.67 &   2.20 &   1.97 &   1.88 &   1.95 &   2.08 \\
\noalign{\smallskip}\hline\noalign{\smallskip}
 IDR14 & Mean $Z$   &  -0.88 &  -0.80 &  -0.04 &  -0.61 &  -1.20 &  -1.61 &  -1.82 &  -2.23 &  -1.82 \\
      &  Median $Z$ &  -0.86 &  -0.74 &  -0.23 &  -0.62 &  -1.18 &  -1.69 &  -1.87 &  -2.38 &  -2.09 \\
       & Std $Z$    &   1.37 &   1.32 &   1.22 &   1.31 &   1.67 &   1.78 &   1.75 &   1.73 &   1.25 \\
       & Med. $LH$  &   0.21 &   0.24 &   0.43 &   0.36 &   0.13 &   0.07 &   0.06 &   0.02 &   0.04 \\
       & Rank        & D     & D     & B     & C     & D     & D     & D     & D     & D    \\
       & $LLH$ Score&   2.83 &   2.65 &   2.00 &   2.49 &   4.06 &   5.19 &   5.67 &   6.85 &   4.69 \\
\noalign{\smallskip}\hline\noalign{\smallskip}
  GK15 & Mean $Z$   &   0.52 &   0.66 &   1.10 &   0.43 &  -0.24 &  -0.86 &  -1.55 &  -2.27 &  -2.25 \\
      &  Median $Z$ &   0.24 &   0.36 &   0.61 &   0.14 &  -0.13 &  -0.74 &  -1.36 &  -2.13 &  -2.40 \\
       & Std $Z$    &   1.60 &   1.68 &   1.93 &   1.53 &   1.53 &   1.66 &   1.74 &   1.64 &   1.31 \\
       & Med. $LH$  &   0.36 &   0.34 &   0.25 &   0.34 &   0.30 &   0.18 &   0.10 &   0.03 &   0.02 \\
       & Rank        & D     & D     & D     & D     & D     & D     & D     & D     & D    \\
       & $LLH$ Score&   2.74 &   3.02 &   4.25 &   2.52 &   2.44 &   3.24 &   4.70 &   6.63 &   6.05 \\
\noalign{\smallskip}\hline\noalign{\smallskip}
ZHAO16 & Mean $Z$   &   0.00 &   0.07 &   0.35 &  -0.23 &  -0.37 &  -0.68 &  -0.94 &  -1.09 &  -1.02 \\
      &  Median $Z$ &   0.02 &   0.09 &   0.42 &  -0.18 &  -0.20 &  -0.77 &  -0.94 &  -1.13 &  -1.12 \\
       & Std $Z$    &   1.17 &   1.16 &   1.19 &   1.09 &   1.32 &   1.39 &   1.46 &   1.36 &   1.10 \\
       & Med. $LH$  &   0.46 &   0.46 &   0.43 &   0.46 &   0.32 &   0.28 &   0.22 &   0.20 &   0.23 \\
       & Rank        & B     & B     & B     & A     & C     & D     & D     & D     & D    \\
       & $LLH$ Score&   1.75 &   1.74 &   1.98 &   1.87 &   2.32 &   2.66 &   3.08 &   3.14 &   2.50 \\
\noalign{\smallskip}\hline\noalign{\smallskip}
BJAN19 & Mean $Z$   &   1.08 &   0.99 &   0.99 &   0.64 &   0.57 &   0.38 &   0.55 &   0.68 &   1.55 \\
      &  Median $Z$ &   1.06 &   0.84 &   0.58 &   0.57 &   0.57 &   0.37 &   0.63 &   0.76 &   0.95 \\
       & Std $Z$    &   1.25 &   1.28 &   1.41 &   1.04 &   1.06 &   1.10 &   1.11 &   1.18 &   2.31 \\
       & Med. $LH$  &   0.20 &   0.29 &   0.31 &   0.42 &   0.38 &   0.34 &   0.34 &   0.32 &   0.21 \\
       & Rank        & D     & D     & D     & C     & C     & B     & C     & D     & D    \\
       & $LLH$ Score&   3.00 &   3.00 &   3.37 &   2.29 &   2.10 &   1.98 &   2.05 &   2.25 &   6.46 \\
\noalign{\smallskip}\hline
\label{reIG} \\
\end{longtable}
}
It is apparent from Table \ref{reIG} that each goodness-of-fit parameters 
show appreciable variation over periods for all the GMPEs. Moreover, no
single GMPE can be considered suitable for the entire range of spectral
periods in this target region. For different periods of interest, different
set of GMPEs are suitable to construct the logic trees in PSHA. For PGA
($0.01$ s), it is suggested to use ZHAO06, IDR08, ZHAO16 and KAN06. For
$0.2$ s, our analysis shows that a combination of BSSA14, BJAN19, AKBO10
and KAN06 may be considered suitable for constructing the logic tree. LLH
score is given more importance in selecting these set of GMPEs as LLH score
is sample size independent. For $1.0$ s, the combination of ASK14, BSSA14,
CY14 and AKBO10 may be considered appropriate for designing of logic tree.
The variation of LLH score at $20$ periods is plotted in Fig. \ref{llhper3}
and the bar diagrams of LLH scores at the three periods of importance
for all the GMPEs are shown in Fig. \ref{llhper4} for easy understanding.
In general, it is observed that for this region NGA-West2 GMPEs
perform better than the NGA-West1 GMPEs for periods $\geq$ $0.2$ s
except for IDR14.
The distribution of normalized residuals,
the best fit normal distribution (continuous curve) with the standard
normal distribution (dashed curve) and the corresponding histograms of
the LH values
are presented in Fig. \ref{lhwch1} for the suitable GMPEs at periods $0.01$ s,
$0.2$ s and $1.0$ s respectively. It may be noted that the median $LH$ values
and the $LLH$ scores behave consistently for all the suitable set of GMPEs in
this target region, too.
\clearpage
\section{Conclusion}
\label{conclu}
With the ever-increasing number of GMPEs with time, the selection of
suitable GMPEs for the construction of logic trees in SHA
for a target region has become a necessity. The construction
of logic trees is a standard practice to address the epistemic 
uncertainties associated with various input parameters. These uncertainties
are relatively larger in regions where limited number of recorded 
ground motion are available. However, this data limitation should not be
used as a basis by the hazard analysts to indulge in subjectivities
while selecting suitable set of GMPEs for carrying out hazard calculations.
\citet{bomm1} pointed out that the selection process ``should neither be
guided by familiarity with certain GMPEs nor by any particular preference
that the analyst may have for a given model''. Rather, the 
appropriateness of GMPEs should be judged by data-driven methods which are
based on sound mathematical background.

To this end, this paper has focused on the selection of the best possible
GMPEs by carrying out a thorough analysis in two geo-tectonic units which
are characterized by distinct geophysical and geo-technical properties. 
For the analysis of suitability of GMPEs, two widely used data-driven
methods proposed by \citet{sche1} and \citet{sche2} are considered. Given
the amount of available data, different suites of suitable GMPEs are 
suggested to be used for different periods of interest in these two target
regions. The results of the ranking of the GMPEs from both the methods are
very consistent. 
The weight factor to be assigned to each selected GMPE
for construction of logic tree is a separate topic, which 
is left for future research. However, in the present case, the weight
factor to each suitable GMPE for a particular target region may be 
decided from the LLH score following the prescription given in
\citet{sche2}, if the readers wish.
The regions under study have witnessed a number of 
significant earthquakes in the recent past and therefore data-driven selection
and ranking of appropriate GMPEs for SHA in these regions
is a necessity. We believe that the results obtained from this research
will help the scientists and engineers entrusted with the task of 
microzonation and earthquake-resistant design of structures in these regions.
The paper is not intended to make any judgement about any particular GMPE
being superior or inferior to any other. Our aim is to provide a 
data-driven selection and ranking of GMPEs that predict the observed 
ground-motion best in these target regions, given the available data at
our disposal.
\section*{Acknowledgements}
Figure \ref{tr} was prepared by using the Generic Mapping Tools
(GMT) software package \citep{wess}, available at
\url{https://www.generic-mapping-tools.org/}. We sincerely acknowledge the
GLOBE Task Team 
\big(\url{http://www.ngdc.noaa.gov/mgg/topo/globe.html})\big) 
for providing the DEM data for preparation of Figure \ref{eqrs}.

The authors are thankful to Dr. R. S. Kankara, Director, 
Shri Rizwan Ali, Scientist - E and Shri Sachin Khupat, 
Scientist - C for their continuous support and encouragement
in the present research. 
The authors thankfully acknowledge two
anonymous referees for their valuable comments which helped in
improving the manuscript.

\section*{Disclosure statement}
The authors have no conflicts of interest to declare that
are relevant to the content of this article.
\section*{Data availability statement}
All the data used in this research are available in public domain and
the links for accessing the data are given inside the text.
\section*{Funding details}
No funding was received for conducting this study.

\bibliographystyle{spbasic}
\bibliography{refs}

\begin{thebibliography}{52}
\providecommand{\natexlab}[1]{#1}
\providecommand{\url}[1]{{#1}}
\providecommand{\urlprefix}{URL }
\expandafter\ifx\csname urlstyle\endcsname\relax
  \providecommand{\doi}[1]{DOI~\discretionary{}{}{}#1}\else
  \providecommand{\doi}{DOI~\discretionary{}{}{}\begingroup
  \urlstyle{rm}\Url}\fi
\providecommand{\eprint}[2][]{\url{#2}}

\bibitem[{Abrahamson and Silva(2008)}]{abra2}
Abrahamson NA, Silva WJ (2008) Summary of the {Abrahamson} and {Silva} {NGA}
  {GRound} {Motion} {Relations}. Earthquake Spectra 24:67--197

\bibitem[{Abrahamson et~al.(2014)Abrahamson, Silva, and Kamai}]{ask14}
Abrahamson NA, Silva WJ, Kamai R (2014) Summary of the {A}{S}{K}14 ground
  motion relation for active crustal regions. Earthquake Spectra
  30(3):1025--1055

\bibitem[{Akkar and Bommer(2010)}]{akbo10}
Akkar S, Bommer J (2010) Empirical equations for the prediction of pga, pgv,
  and spectral accelerations in {E}urope, the {M}editerranean region, and the
  {M}iddle {E}ast. Seismological Research Letters 81(2):195--206

\bibitem[{Anbazhagan et~al.(2015)Anbazhagan, Bajaj, Moustafa, and
  Al-Arifi}]{anba3}
Anbazhagan P, Bajaj K, Moustafa SSR, Al-Arifi NSN (2015) Maximum magnitude
  estimation considering the regional rupture character. Journal of Seismology
  19:695--719

\bibitem[{Arango et~al.(2012)Arango, Free, Lubkowski, Pappin, Musson, Jones,
  and Hodge}]{aran1}
Arango MC, Free MW, Lubkowski ZA, Pappin JW, Musson RMW, Jones G, Hodge E
  (2012) Comparing predicted and observed ground motions from {U}{K}
  earthquakes. In: 15th {W}orld {C}onf. on {E}arthquake {E}ngineering

\bibitem[{B.~Naresh and Mishra(2021)}]{nar21}
B~Naresh KV, Mishra LK (2021) The seismotectonic setting of indo-gangetic plain
  and its importance. In: Seismic Hazards and Risk: Select Proceedings of 7th
  ICRAGEE 2020 (Lecture Notes in Civil Engineering), vol 116, pp 187--196

\bibitem[{Bajaj and Anbazhagan(2019)}]{bjan19}
Bajaj K, Anbazhagan P (2019) Regional stochastic {G}{M}{P}{E} with available
  recorded data for active region – application to the {H}imalayan region.
  Soil Dynamics and Earthquake Engineering 126:105825

\bibitem[{Beauval et~al.(2012)Beauval, Tasan, Laurendeau, Delavaud, Cotton,
  Gueguen, and Kuehn}]{beau1}
Beauval C, Tasan H, Laurendeau A, Delavaud E, Cotton F, Gueguen P, Kuehn N
  (2012) On the testing of ground-motion prediction equations against small
  magnitude data. Bull Seism Soc Am 102:1994--2007

\bibitem[{Bommer and Scherbaum(2008)}]{bomm2}
Bommer JJ, Scherbaum F (2008) The use and misuse of logic-trees in
  {P}{S}{H}{A}. Earthquake Spectra 24:997--1009

\bibitem[{Bommer et~al.(2010)Bommer, Douglas, Scherbaum, Cotton, Bungum, and
  Fah}]{bomm1}
Bommer JJ, Douglas J, Scherbaum F, Cotton F, Bungum H, Fah D (2010) On the
  selection of ground-motion prediction equations for seismic hazard analysis.
  Seismological Research Letters 81:783--793

\bibitem[{Boore et~al.(2014)Boore, Stewart, Seyhan, and Atkinson}]{bssa14}
Boore DM, Stewart JP, Seyhan E, Atkinson GM (2014) {N}{G}{A}-{W}est2 equations
  for predicting pga, pgv, and 5$\%$ damped psa for shallow crustal
  earthquakes. Earthquake Spectra 30(3):1057 -- 1085

\bibitem[{Boore and Atkinson(2008)}]{boore1}
Boore MD, Atkinson GM (2008) Ground-motion prediction equations for the average
  horizontal component of {PGA}, {PGV}, and 5$\%$-damped {PSA} at spectral
  periods between 0.01 s and 10.0s. Earthquake Spectra 24:99--138

\bibitem[{Bykova(2016)}]{byko1}
Bykova VV (2016) On the selection of ground-motion prediction equations during
  the assessment of seismic hazard in stable continental regions. Seismic
  Instruments 52:135--143

\bibitem[{Campbell and Bozorgnia(2008)}]{cam4}
Campbell KW, Bozorgnia Y (2008) N{G}{A} ground motion model for the geometric
  mean horizontal component of {PGA}, {PGV}, {PGD} and 5$\%$ damped linear
  elastic response spectra for periods ranging from 0.01 to 10s. Earthquake
  Spectra 24:139--171

\bibitem[{Campbell and Bozorgnia(2014)}]{cb14}
Campbell KW, Bozorgnia Y (2014) N{G}{A}-{W}est2 ground motion model for the
  average horizontal components of pga, pgv, and 5$\%$ damped linear
  acceleration response spectra. Earthquake Spectra 30(3):1087 -- 1115

\bibitem[{Chiou and Youngs(2008)}]{chi1}
Chiou B, Youngs RR (2008) An {NGA} model for the average horizontal component
  of peak ground motion and response spectra. Earthquake spectra 24:173--215

\bibitem[{Chiou and Youngs(2014)}]{cy14}
Chiou BSJ, Youngs RR (2014) Update of the chiou and youngs nga model for the
  average horizontal component of peak ground motion and response spectra.
  Earthquake Spectra 30(3):1117 -- 1153

\bibitem[{Cotton et~al.(2006)Cotton, Scherbaum, Bommer, and Bungum}]{cott1}
Cotton F, Scherbaum F, Bommer JJ, Bungum H (2006) Criteria for selecting and
  adjusting ground-motion models for specific target applications: Applications
  to central europe and rock sites. Journal of Seismology 10:137--156

\bibitem[{Cover and Thomas(2006)}]{cove1}
Cover TM, Thomas JA (2006) Elements of {I}nformation {T}heory, {S}econd ed.,
  {W}iley, {H}oboken, {N}ew {J}ersey

\bibitem[{Dasgupta et~al.(2000)Dasgupta, Pande, Ganguly, Iqbal, Sanyal,
  Venaktraman, Dasgupta, Sural, Harendranath, Mazumdar, Sanyal, Roy, Das,
  Misra, and Gupta}]{dasg1}
Dasgupta S, Pande P, Ganguly D, Iqbal Z, Sanyal K, Venaktraman NV, Dasgupta S,
  Sural B, Harendranath L, Mazumdar K, Sanyal S, Roy A, Das LK, Misra PS, Gupta
  H (2000) Seismotectonic {Atlas} of {India} and {Its} {Environs}. Geological
  Survey of India:Calcutta

\bibitem[{Delavaud et~al.(2009)Delavaud, Scherbaum, Kuehn, and
  Riggelsen}]{dela1}
Delavaud E, Scherbaum F, Kuehn N, Riggelsen C (2009) Information- theoretic
  selection of ground-motion prediction equations for seismic hazard analysis:
  An applicability study using {C}alifornia data. Bull Seism Soc Am
  99:3248--3263

\bibitem[{Douglas(2007)}]{doug2}
Douglas J (2007) On the regional dependence of earthquake response spectra.
  ISET {J}ournal of Earthquake Technology 44:71--99

\bibitem[{Douglas(2022)}]{doug1}
Douglas J (2022) Ground motion prediction equations 1964-2021.
  \url{http://www.gmpe.org.uk}

\bibitem[{Edwards(1992)}]{edwa1}
Edwards AWF (1992) Likelihood, {E}xpanded ed., {J}ohns {H}opkins {U}niv.
  {P}ress

\bibitem[{Graizer and Kalkan(2016)}]{gk15}
Graizer V, Kalkan E (2016) Summary of {G}{K}15 ground-motion prediction
  equation for predicting {P}{G}{A} and 5$\%$ damped {S}{A} from shallow
  crustal continental earthquakes. Bulletin of Seismological Society of America
  106(2):687--707

\bibitem[{Gupta(2006)}]{gup06}
Gupta ID (2006) Delineation of probable seismic sources in india and
  neighbourhood. Soil Dynamics and Earthquake Engineering 26:766--790

\bibitem[{Gupta and Trifunac(2018)}]{gupta5}
Gupta ID, Trifunac MD (2018) Empirical scaling relations for pseudo relative
  veocity spectra in western {Himalaya} and northeastern {India}. Soil Dynamics
  and Earhquake engineering 106:70--89

\bibitem[{Hintersberger et~al.(2007)Hintersberger, Scherbaum, and
  Hainzl}]{hint1}
Hintersberger E, Scherbaum F, Hainzl S (2007) Update of likelihood-based
  ground-motion model selection for seismic hazard analysis in western central
  {E}urope. Bull Earth Engg 5:1--16

\bibitem[{Idriss(2008)}]{idr08}
Idriss IM (2008) An {N}{G}{A} empirical model for estimating the horizontal
  spectral values generated by shallow crustal earthquakes. Earthquake Spectra
  24(1):217--242

\bibitem[{Idriss(2014)}]{idr14}
Idriss IM (2014) An {N}{G}{A}-{W}est2 empirical model for estimating the
  horizontal spectral values generated by shallow crustal earthquakes.
  Earthquake Spectra 30(3):1155 -- 1177

\bibitem[{Kale and Akkar(2013)}]{kale1}
Kale O, Akkar S (2013) A new procedure for selecting and ranking ground-motion
  prediction equations (gmpes): The {E}uclidean distance-based ranking
  ({E}{D}{R}) method. Bull Seism Soc Am 103:1069--1084

\bibitem[{Kanno et~al.(2006)Kanno, Narita, Morikawa, Fujiwara, and
  Fukushima}]{kann1}
Kanno T, Narita A, Morikawa N, Fujiwara H, Fukushima Y (2006) A new attenuation
  releation for strong ground motion in {Japan} based on recorded data.
  Bulletin of the seismological Society of America 96:879--897

\bibitem[{Kayal(2010)}]{kaya10}
Kayal JR (2010) Himalayan tectonic model and the great earthquakes: an
  appraisal. Geomatics, Natural Hazards and Risk 1(1):51--67

\bibitem[{Kayal(2014)}]{kaya14}
Kayal JR (2014) Seismotectonics of the great and large earthquakes in himalaya.
  Current Science 106(2):188--197

\bibitem[{Kayal et~al.(2022)Kayal, Baruah, Hazarika, and Das}]{kaya22}
Kayal JR, Baruah S, Hazarika D, Das A (2022) Recent large and strong
  earthquakes in the eastern himalayas: An appraisal on seismotectonic model.
  Geological Journal 57(12):4929--4938

\bibitem[{Kowsari et~al.(2019)Kowsari, Halldorsson, Hrafnkelsson, and
  Jonsson}]{kowa1}
Kowsari M, Halldorsson B, Hrafnkelsson B, Jonsson S (2019) Selection of
  earthquake ground motion models using the deviance information criterion.
  Soil Dynam Earthq Engg 117:288--299

\bibitem[{Kulkarni et~al.(1984)Kulkarni, Youngs, and Coppersmith}]{kulk1}
Kulkarni RB, Youngs RR, Coppersmith KJ (1984) Assessment of confidence
  intervals for results of seismic hazard analysis. In: 8th {W}orld {C}onf. on
  {E}arthquake {E}ngineering, pp 263--270

\bibitem[{Kumar et~al.(2016)Kumar, Singh, Mitra, Priestly, and Dayal}]{ajay1}
Kumar A, Singh S, Mitra S, Priestly KF, Dayal S (2016) The 2015 {A}pril 25
  {G}orkha ({N}epal) earthquake and its aftershocks: implications for lateral
  heterogeneity of the {M}ain {H}imalayan {T}hrust. Geophysical Journal
  International 208(2):992--1008

\bibitem[{Nath et~al.(2012)Nath, Thingbaijam, Maity, and Nayak}]{nath3}
Nath SK, Thingbaijam K, Maity S, Nayak A (2012) Ground-motion predictions in
  {Shilong} region, {North-East} {India}. Journal of Seismology 16:475--488

\bibitem[{Nayak and Sitharam(2019)}]{mona1}
Nayak M, Sitharam TG (2019) Estimation and spatial mapping of seismicity
  parameters in western {H}imalaya, central {H}imalaya and {I}ndo-{G}angetic
  plain. J Earth Syst Sci 128:45

\bibitem[{Nh and Kumar(2020)}]{hari1}
Nh H, Kumar A (2020) Ground motion prediction equation for north {I}ndia,
  applicabale for different site classes. Soil Dynamics and Earhquake
  engineering 139:106425

\bibitem[{Scherbaum et~al.(2004)Scherbaum, Cotton, and Smit}]{sche1}
Scherbaum F, Cotton F, Smit P (2004) On the use of response spectral-reference
  data for the selection and ranking of ground-motion models for seismic-hazard
  analysis in regions of moderate seismicity: The case of rock motion. Bull
  Seism Soc Am 94:2164--2185

\bibitem[{Scherbaum et~al.(2009)Scherbaum, Delavaud, and Riggelsen}]{sche2}
Scherbaum F, Delavaud E, Riggelsen C (2009) Model selection in seismic hazard
  analysis: an information- theoretic perspective. Bull Seism Soc Am
  99:3234--3247

\bibitem[{Sharma et~al.(2009)Sharma, Douglas, Bungum, and Kotadia}]{shar1}
Sharma ML, Douglas J, Bungum H, Kotadia J (2009) Groung motion prediction
  equations based on data from the {Himalayan} and {Zagros} region. Journal of
  Earthquaje Engineering 13:1191--1210

\bibitem[{Singh et~al.(2017)Singh, Srinagesh, Srinivas, Arroyo, amd
  R.~K.~Chadha, Suresh, and Suresh}]{singh1}
Singh SK, Srinagesh D, Srinivas D, Arroyo D, amd R~K~Chadha XPC, Suresh G,
  Suresh G (2017) Strong {G}round {M}otion in the {I}ndo-{G}angetic {P}lains
  during the 2015 {G}orkha, {N}epal, {E}arthquake {S}equence and {I}ts
  {P}rediction during {F}uture {E}arthquakes. Bulletin of the Seismological
  Society of America 107(3):1293--1306

\bibitem[{Sinha and Selvan(2022)}]{sum22}
Sinha S, Selvan S (2022) An improved probabilistic seismic hazard assessment of
  {T}ripura, {I}ndia. Pure and Applied Geophysics 179:4371--4393

\bibitem[{Slejko and Rebez(2002)}]{slej1}
Slejko D, Rebez A (2002) Probabilistic seismic hazard assessment and
  deterministic ground shaking scenarios for {V}ittorio {V}eneto
  ({N}.{E}.{I}taly) volume = {43}, journal = {Boll. Geof. Teor. Appl.}, pp
  263--280

\bibitem[{Stafford et~al.(2008)Stafford, Strasser, and Bommer}]{staf1}
Stafford PJ, Strasser FO, Bommer JJ (2008) An evaluation of the applicability
  of the nga models to ground-motion prediction in the euro-mediterranean
  region. Bulletin of Earthquake Engineering 6:149--177

\bibitem[{Strasser et~al.(2009)Strasser, Abrahamson, and Bommer}]{stra1}
Strasser FO, Abrahamson NA, Bommer JJ (2009) Sigma: Issues, insights and
  challenges. Seismological Research Letters 80:40--56

\bibitem[{Wessel et~al.(2019)Wessel, Luis, Uieda, Scharroo, Wobbe, Smith, and
  Tian}]{wess}
Wessel P, Luis FJ, Uieda L, Scharroo R, Wobbe F, Smith WHF, Tian D (2019) The
  {Generic} {Mapping} {Tools} version 6 pp 5556--5564

\bibitem[{Zhao et~al.(2006)Zhao, Zhang, Asano, Ohno, Oouchi, Takahashi, Ogawa,
  Irikura, Thio, Somerville, Fukushima, and Fukushima}]{zhao06}
Zhao JX, Zhang J, Asano A, Ohno Y, Oouchi T, Takahashi T, Ogawa H, Irikura K,
  Thio HK, Somerville PG, Fukushima Y, Fukushima Y (2006) Attentuation
  relations of strong ground motion in {J}apan using site classification based
  on predominant period. Bulletin of the Seismological Society of America
  96(3):898 -- 913

\bibitem[{Zhao et~al.(2016)Zhao, Zhou, Gao, Zhang, Zhou, Lu, and
  Rhoades}]{zhao16c}
Zhao JX, Zhou SL, Gao PJ, Zhang YB, Zhou J, Lu M, Rhoades DA (2016)
  Ground-motion prediction equations for shallow crustal and upper mantle
  earthquakes in {J}apan using site class and simple geometric attenuation
  functions. Bulletin of the Seismological Society of America 106:1552 -- 1569

\end{thebibliography}

\end{document}